\documentclass[useAMS,usegraphicx,usenatbib,11pt]{mn2e}
\usepackage{subfigure}
\usepackage[fleqn]{amsmath}
\usepackage{amssymb}
\usepackage[english]{babel}  
\usepackage{array}
\usepackage{graphicx}
\usepackage{multicol}
\usepackage{times}

\title[A code for the evolution of star clusters]{A prescription and fast code for the long-term evolution of star clusters}
\author[P.~E.~R.~Alexander \& M. Gieles]{Poul~E.~R.~Alexander$^{1}$\thanks{e-mail: pera@ast.cam.ac.uk} and Mark~Gieles$^{1}$\thanks{e-mail: mgieles@ast.cam.ac.uk} \\ $^1$Institute of Astronomy, University of Cambridge, Madingley Road, Cambridge, CB3 0HA, UK}
\date{Accepted 2012 February 29; Received 2012 February 24; in original form: 2012 December 21}

\begin{document} 

\newcommand{\rhrj}{\mathcal{R}} 
\newcommand{\rhrjr}{\mathcal{R}_{1}}   
\newcommand{\rvrj}{\mathcal{R}} 
\newcommand{\rvrjr}{\mathcal{R}_1} 
\newcommand{\rhrjo}{\mathcal{R}_0}
\newcommand{\rvrjo}{\mathcal{R}_0} 
\newcommand{\dr}{{\rm d}}
\newcommand{\erf}{{\rm erf}}
\newcommand{\erfi}{{\rm erfi}}
\newcommand{\rh}{r_{\rm h}}
\newcommand{\rv}{r_{\rm v}}
\newcommand{\rj}{r_{\rm J}}

\newcommand{\sovast}{SvA} 
\newcommand{\nat}{Nature} 
\newcommand{\pasj}{PASJ} 
\newcommand{\pasp}{PASP}
\newcommand{\apss}{Ap\&SS}
\newcommand{\jcp}{Journal of Chemical Physics} 
\newcommand{\ApJL}{Ap. J. Letters}
\newcommand{\ApJ}{Ap.J.}
\newcommand{\apjs}{Ap.J. Supplements}
\newcommand{\PRL}{Phys. Rev. Letters}
\newcommand{\apj}{Ap.J.}
\newcommand{\apjl}{Ap. J. Letters}
\newcommand{\aj}  {AJ}
\newcommand{\aap}{A\&A}
\newcommand{\MNRAS}{MNRAS}
\newcommand{\mnras}{MNRAS}
 
\maketitle

\begin{abstract}
We introduce the star cluster evolution code \textsc{Evolve Me A Cluster of StarS (EMACSS)}, a simple yet physically motivated computational model that describes the evolution of some fundamental properties of star clusters in static tidal fields. We base our prescription upon the flow of energy within the cluster, which is a constant fraction of the total energy per half-mass relaxation time. According to H{\'e}non's predictions, this flow is independent of the precise mechanisms for energy production within the core, and therefore does not require a complete description of the many-body interactions therein. For a cluster of equal-mass stars, we thence use dynamical theory and analytic descriptions of escape mechanisms to construct a series of coupled differential equations expressing the time-evolution of cluster mass and radius. These equations are numerically solved using a 4th order Runge-Kutta integration kernel, and the results bench-marked against a database of direct $N$-body simulations. We use simulations containing a modest initial number of stars ($1024\le N\le 65536$), and point-mass tidal fields of various strengths. Our prescription is publicly available, and reproduces the $N$-body results to within $\sim10\%$ accuracy for the entire post-collapse evolution of star clusters.
\end{abstract}
\begin{keywords}
stellar dynamics: methods -- galaxies: star clusters -- globular clusters: general -- methods: $N$-body simulations --methods: Numerical
\end{keywords}

\section{Introduction}
The evolution of a star cluster is driven by a combination of relaxation (\citealt{Amb1938,Chandra1942,King1958}), binary interactions (\citealt{Heggie1975}), stellar evolution and stellar encounters (\citealt{Hut1992}). Furthermore, for a cluster located within the tidal field of a galaxy, effects resulting from cluster's interaction with the tidal field can be important (\citealt{Henon1960,Lee1987}), with the consequences of a limiting `Jacobi' radius, and a shortened total lifetime (\citealt{Baumgardt2003}). The result is a complex system, in which the simultaneous modelling of several properties (e.g. mass, half-mass radius, density profile) is equivocal. 

As a result of this complexity, dynamical simulations are an appealing manner through which to study the evolution of star clusters. Significant success has been achieved through direct $N$-body integrations (see \citealt{Aarseth1974,Makino1996,Spurzem1999}), albeit at the expense of high computational cost. Likewise, faster schemes (Monte Carlo schemes: \citealt{Henon1975,Giersz1998}, solving the Fokker-Planck equation: \citealt{Cohn1979}, Gas models: \citealt{Larson1970}) have each obtained significant success modelling high $N$-systems, although with a reduction in versatility owing to the assumptions required for each (\citealt{Heggie2003}). Of particular relevance to this study, the Fokker-Planck equation represents one successful and explicit formula for modelling the evolution of \emph{the distribution function} of a many-body system, in a manner similar to our intended prescription. However, this formula generally does not unambiguously express the dynamical properties for which we intend our prescription (\citealt{Zwart1998}). 
 
Qualitatively, the long-term evolution of a star cluster has been characterised into three phases; core collapse, core bounce and expansion, and tidally limited contraction. The first stage (core collapse) has historically been the most studied, and results from the diffusion of kinetic energy from core stars outward through relaxation. These stars, upon loosing kinetic energy, move inward and thus experience a deeper potential. As a result, these inward-moving stars accelerate, eventually leading to a net increase in core kinetic energy (a consequence encapsulated by the concept of negative heat capacity in gravothermal systems, \citealt{LB-Wood1968}). However, the increasing kinetic energy of core stars will ultimately enhance the rate at which relaxation diffuses energy from the core, which in turn accelerates the process of collapse.

Core collapse is eventually halted once a source of energy becomes viable (see \citealt{Statler1987,Giersz1994}), and energy removed from the core can be replaced without the need for further contraction. At this point, the core `bounces' outward owing to excess energy released during collapse (\citealt{Inagaki1983}), before (for an isolated or compact cluster) entering an expansion phase wherein outflowing energy from the core inflates the cluster (\citealt{Henon1965-1,Lightman1978}). During this process, the rate of energy production in the core comes into balance with the flow of energy that is required for the global evolution. A final contraction stage occurs once the system has expanded sufficiently that the evaporation of stars over the Jacobi radius has become the dominant evolutionary consequence (\citealt{Gieles2008}), and the evolution of the system is defined by the rate at which stars are lost to the tidal field.

\citet{Gieles2011} have proposed a formalism through which the expansion and tidally limited phases can be linked, and thus a complete description of the life cycle obtained. The two phases therefore constitute extreme cases, defining asymptotic behaviour in the evolution of clusters. Making the assumption of self-similar evolution, \citet{Henon1965-1} showed that, for expanding clusters (without escaping stars) the half-mass relaxation time scales linearly with time ($t_{\textrm{rh}} \propto t$), and therefore the half-mass radius scales with $t^{2/3}$. By similar methodology, \citet{Henon1961} predicted that if a cluster is limited by it's Jacobi radius, the mass decreases linearly in time, and the half-mass radius will scale linearly with the Jacobi radius. Hence, the half-mass radius will scale with $(t_{\textrm{ev}}-t)^{1/3}$ where $t_{\textrm{ev}}$ is the total lifetime, or evaporation time of a cluster. These two power-laws constitute extremes of the evolution, with much of a cluster's life cycle forming a transition between these two phases.

For the two interpretations quoted above, H{\'e}non considered evolution to occur in a \emph{self-similar} fashion (described in \citet{Henon1961} as being \emph{homologous}); the shape of the density profile remains constant, with an isothermal cusp and finite truncation at the Jacobi radius. Accordingly, if a cluster evolves self-similarly, the dynamical effects are limited to a variation in overall scale, and a corresponding gradual reduction of total mass. Although this simplifying assumption is not valid in extreme cases (prior to the establishment of balanced evolution, or when close to final dissolution and containing a relatively low number of stars), we initially employ this interpretation before exploring the effects of none self-similar evolution.

The goal of this paper is to present the initial stage in the development of a dynamical prescription encompassing both the expansion and tidal contraction phases. To this end, our initial prescription is strongly simplified; we consider only the dynamical effects stemming from the interaction of equal-mass stars without internal evolution, and thereupon eliminate the complications of mass segregation. Although these effects are important for realistic globular clusters (\citealt{Spitzer1969,Spitzer1975}), such effects serve here mainly to complicate the dynamics we seek to study, and anyway do not necessarily introduce effects not otherwise represented in an equal mass model (\citealt{LB1980}). We allow our models no primordial binary content, and model our stars as point particles so as to eliminate perturbations caused by direct stellar collisions. Our approximation for the tidal field into which we immerse the cluster is also simplified, as we model the tidal field to be that of a point mass galaxy located well outside the cluster. The ensuing Jacobi radius is regarded as a spherical surface, although we do take into account the effect of preferential trajectories for escape (see section~\ref{s:t4}). The model is constructed in the form of an efficient \textsc{C++} code\footnote{available at https://github.com/emacss}, which we calibrate against $N$-body simulations.

In section~\ref{s:EvSC} we discuss the mechanics and physical details of escape considered by our prescription, which is itself discussed in section~\ref{s:4}. Following this, in section~\ref{s:3} we discuss our $N$-body simulations, against which our prescription is calibrated and verified in section~\ref{s:res}. Finally, we consider the success and shortcomings of our prescription in section~\ref{s:6}, and outline the future physics that is to be incorporated into our model. 

\section{Evolution of Star Clusters}
\label{s:EvSC}

The dynamical evolution of a star cluster is a process primarily driven by the radial diffusion of energy from the `hot' (energetic) core (\citealt{VH1957,Henon1961,Larson1970}). Thus, the system's life cycle is spent striving to establish equipartition. Equipartition remains elusive however, on account of the inherent negative heat capacity of gravothermal systems, with the result that the flow of energy continually brings the cluster further out of equilibrium. This means that the energy `produced' in the core\footnote{Principally by  dynamical mechanisms such as the formation and hardening of binaries during three body encounters (\citealt{Heggie1975}), although other options such as stellar evolution are viable (\citealt{Gieles2010}).} is continually transferred by relaxation outward into the halo of the cluster, where the injection of energy drives the dynamical effects we seek to study.

We shall examine star clusters both within a tidal field, and isolated from external tidal influences. Although such isolated clusters do not naturally exist, models of isolated clusters describe the expansion phase experienced in the early evolution of (realistic) tidally limited clusters\footnote{To quote from \citet{Aarseth1998},``much may be learned by the study of more tractable, idealised models, provided that the goal of understanding the behaviour of real clusters is always kept in mind.''} (\citealt{Aarseth1971}).

We begin by taking the familiar expression for the virial radius of a star cluster, $r = -GM^2/2W$, where $W$ is potential energy, $G$ is the gravitational constant and $M$ the total mass. Assuming that the system remains in virial equilibrium throughout its evolution such that the total energy $E = W/2$, we can express this total energy as
\begin{align}
 E = -\frac{GN^2\bar{m}^2}{4r},
\label{eq:EEv}
\end{align}
where $M = N\bar{m}$, with $N$ the number of stars and $\bar{m}$ the mean mass of stars. Alternatively, we can express energy in terms of the half-mass radius $r_{\textrm{h}}$, such that
\begin{align}
 E = -\kappa \frac{GN^2\bar{m}^2}{r_{\textrm{h}}}, 
\label{eq:EE}
\end{align}
in which $\kappa$ is a form factor dependent upon the density profile whereby  $r_{\rm h} = 4 \kappa r$. However, here we shall make no distinction, i.e. we assume $\kappa = 1/4$.

 From these two expressions, we can begin to parametrise the implications of the outward diffusion of energy by making two initial assumptions. Firstly, we assume that the flux of energy passing through any shell is equal to the change in energy inside that shell. Hence, if a cluster has achieved balanced evolution, the flux of energy passing through a spherical shell located at any given radius is equivalent to that released form the core. Secondly, we assume that the fraction of the total energy passing through this shell is constant per relaxation time, as the system is restored to energetic equilibrium over this timescale. Thus we find
\begin{align}
\frac{\dot{E}}{|E|} = \frac{\zeta}{t_{\textrm{rh}}},\label{eq:EdotE1}
\end{align}
where $\zeta$ is a dimensionless constant and $t_{\textrm{rh}}$ is the half-mass relaxation time. In accordance with the derivation presented by \citet{Spitzer1971}, we define $t_{\textrm{rh}}$ as
\begin{align}
t_{\textrm{rh}} = 0.138\frac{N^{1/2}r_{\textrm{h}}^{3/2}}{\sqrt{\bar{m}G} \ln(\gamma N) }, \label{eq:relax}
\end{align}
in which $\ln(\gamma N)$ is the Coulomb logarithm with $\gamma \approx 0.11$ for equal mass clusters (see \citealt{Giersz1994}). Using equation~(\ref{eq:EEv}) and noting that $\bar{m}$ is constant for an equal-mass system, it is evident that any change in energy will depend only upon $N$ and $r$. Hence
\begin{align}
\frac{\dot{E}}{|E|} = -2 \frac{\dot{N}}{N} + \frac{\dot{r}}{r}, \label{eq:EdotEv}
\end{align}
 while equation~(\ref{eq:EE}) would result in an equivalent expression in which $\dot{\kappa}/\kappa$ would be included. However, as we have fixed $\kappa = 1/4$, we approximate the evolution of both virial and half-mass radius by equation~(\ref{eq:EdotEv}).

Equation~(\ref{eq:EdotEv}) demonstrates that a change of energy can have two principle dynamical effects; either stars are lost, carrying energy away out of system, and/or the radius will vary. It is appropriate at this stage is to parametrise the change in $N$ and $r$ per $t_{\textrm{rh}}$ in terms of dimensionless escape and expansion rates $\xi$ and $\mu$ (\citealt{Goodman1984,Baumgardt2002}). From the definitions therein we therefore take,
\begin{align}
\frac{\dot{N}}{N} &= -\frac{\xi}{t_{\textrm{rh}}}, \label{eq:NdotN}   
\end{align}
and 
\begin{align}
\frac{\dot{r}}{r} = \frac{\mu}{t_{\textrm{rh}}}. \label{eq:rdotr}
\end{align}
If only one phase of evolution is modeled, both $\xi$ and $\mu$ can be considered to be constant (ibid.). By contrast, since we attempt to model both expansion and (in a tidal field) contraction concurrently, we must allow $\xi$ and $\mu$ to vary throughout the life cycle. By combining equations~(\ref{eq:EdotE1}), (\ref{eq:NdotN}) and~(\ref{eq:rdotr}), we find $\zeta$ is related to $\xi$ and $\mu$ by
\begin{align}
\zeta = \mu + 2\xi, \label{eq:relate}
\end{align}
at all times throughout the life cycle. We hence show that the time-evolution of $N$ and $r$ will be defined by a set of differential equations expressing $\dot{E}$, $\dot{N}$ and $\dot{r}$ in terms of $\zeta$, $N$ and $r$. We now examine these equations for a cluster in isolation (thereupon only experiencing it's expansion phase, section~\ref{s:t1}), or a tidal field (and hence experiencing both expansion and contraction, sections~\ref{s:t3} and~\ref{s:t4}), and combine these in section~\ref{s:4}. 

\subsection{Isolated clusters} 
\label{s:t1} \label{s:t2}
Both expansion and mass-loss occur in isolated clusters (\citealt{Baumgardt2002}) meaning that equations~(\ref{eq:NdotN}) and~(\ref{eq:rdotr}) must be solved simultaneously. Dividing equation~(\ref{eq:rdotr}) by equation~(\ref{eq:NdotN}) and eliminating $\mu$ with equation~(\ref{eq:relate}), we obtain the differential equation, 
\begin{align}
\frac{\textrm{d}r}{\textrm{d}N} = \frac{r}{N}\left(2-\frac{\zeta}{\xi}\right). \label{eq:dNdr}
\end{align}
It has been shown (\citealt{Baumgardt2002} and references contained) that an isolated cluster will expand nearly (although not entirely) homologously throughout it's lifetime, loosing a roughly constant fraction of it's stars per relaxation time. As such, it is permissible to attempt a solution with constant $\xi$. Making this assumption and writing the constant $\xi$ of isolated clusters as $\xi_1$, equation~(\ref{eq:dNdr}) is separable and (following integration) yields

\begin{align} 
\frac{r}{r_0} = \left[\frac{N}{N_0}\right]^{2-\frac{\zeta}{\xi_1}},
\label{eq:Nasrv}
\end{align}
where $N_0$ and $r_0$ are scale constants, usually considered to correspond to the cluster at the time of core collapse. As we have assumed $r = r_{\textrm{h}}$, we can eliminate the $r_{\textrm{h}}$ dependence in relaxation time such that
\begin{align} 
t_{\textrm{rh}} = t_{\textrm{rh,0}} \left(\frac{N}{N_0}\right)^{-\frac{1}{\nu}}\left(\frac{\ln(\gamma N_0)}{\ln(\gamma N)}\right),\label{eq:trhscal2}
\end{align}
where,
\begin{align}
\nu = \frac{2\xi_1}{3\zeta-7\xi_1}. \label{eq:nu}
\end{align}
 If we now make the assumption that the Coulomb logarithm is constant, we can explicitly compute the evolution of $N$ and $r$ by applying equation~(\ref{eq:trhscal2}) to~(\ref{eq:NdotN}) and solving. Thus,
\begin{align}
\frac{N}{N_0} = \left(1+\frac{\xi_1}{\nu t_{\textrm{rh,0}}}\left(t-t_0\right)\right)^{-\nu} \label{eq:Baum1},
\end{align}
while, by a similar method,
\begin{align}
\frac{r}{r_0} = \left(1+\frac{\xi_1}{\nu t_{\textrm{rh,0}}}\left(t-t_0\right)\right)^{\frac{2+\nu}{3}}, \label{eq:Baum2}
\end{align}  
Equations~(\ref{eq:Baum1}) and~(\ref{eq:Baum2}) demonstrate a power-law scaling, and are dependent upon the time at which evolution begins, $t_0$. Choosing this time as,
\begin{align}
t_0 = \frac{\nu}{\xi}t_{\textrm{rh,0}}, \label{eq:t0} 
\end{align}
we obtain, 
\begin{align}
N &= N_0\left(\frac{t}{t_0}\right)^{-\nu}, \label{eq:Nv1}\\
r &= r_0\left(\frac{t}{t_0}\right)^{(2+\nu)/3}, \label{eq:rv1}
\end{align}
which are identical to the forms suggested by \citet{Goodman1984}. Our chosen definition of $t_0$, equation~(\ref{eq:t0}) is here used as a numerical tool to simplify equations~(\ref{eq:Baum1}) and~(\ref{eq:Baum2}), although implicitly assumes that the solution of $N$ and $r$ will pass through the origin at $t=0$. This point ($t_0$) corresponds to the moment at which a cluster enters balanced evolution, and roughly to the core collapse time of the system.

\begin{figure}
\vspace{10mm}
 \centering
   \includegraphics[width=80mm,height=110mm]{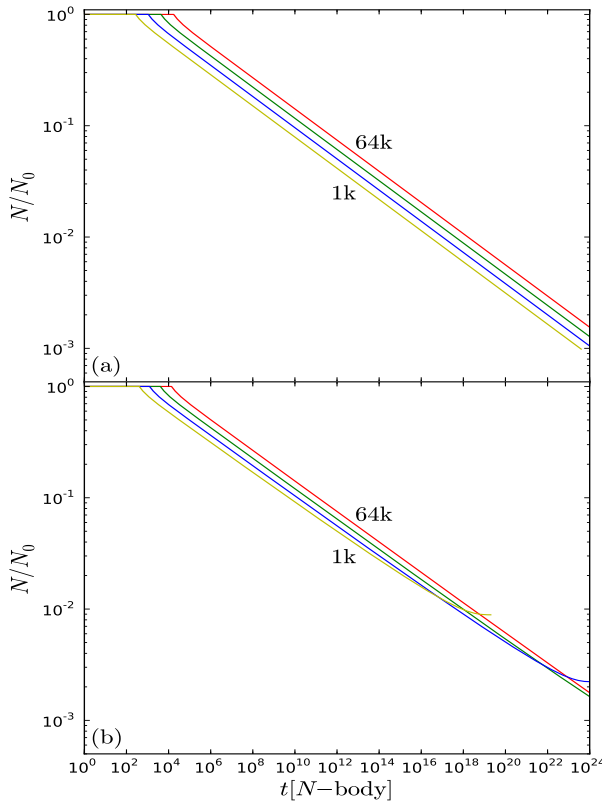}
   \label{fig:ntfu}
 \caption{The effect of the two evolutionary models described in section~\ref{s:t1}, produced by integration of the equations~(\ref{eq:NdotN}) and~(\ref{eq:rdotr}). (a) Constant $\xi_1$ and invariant Coulomb logarithm. (b) Constant $\xi_1$ and $N$-dependent Coulomb logarithm. The multiple lines denote clusters with $N_0$ determined by successive factors of four, between 1024 (labeled 1k) and 65536 (labeled 64k). The different points at which mass loss begins are a physical effect of the variation of $t_{\textrm{rh}}$ with $N$, as for a Plummer Model the core collapse time is approximated by a constant number (for this illustration, 20) of initial $t_{\textrm{rh}}$. For the illustrations, $\zeta = 0.1$ and $\xi_1 = 0.01$.}
 \label{f:notidecomp}
\vspace{10mm}
\end{figure}

Equations~(\ref{eq:Nv1}) and~(\ref{eq:rv1}) imply that the curves $N(t)$ and $r(t)$ are parallel in a log-log figure for a cluster of any $N_0$ (see figure~\ref{f:notidecomp}(a)). Inclusion of the $N$ dependence of the Coulomb logarithm (numerically, since the logarithmic terms retained in equation~(\ref{eq:Baum1}) prohibit integration of equation~(\ref{eq:rdotr})) modifies this evolution to allow a limited convergence of the evolutionary tracks, which is especially apparent at late times (figure~\ref{f:notidecomp}(b)). 

However, \citet{Baumgardt2002} show (via $N$-body simulations) that there is in fact a crossing of the evolutionary tracks (i.e. $\xi \equiv \xi(N)$, with larger $\xi$ for larger $N$). This effect is most likely a result of a positive correlation between core density and $N$ for a star cluster (\citealt{Heggie2003}). Stars escaping from an isolated cluster will by necessity escape to infinity, as there is no possibility of escape via translation over a tidal boundary. It follows that these stars will have positive kinetic energy, obtained by acceleration during the close encounters of two or more bodies in the core (\citealt{Henon1960}). The probability of an encounter leading to the escape of a star will be a function of  number density and velocity dispersion of the stars, with most escapers predicted to come from high density regions (namely, the central core). Thus, it is logical to suggest that an increased core density will inevitably lead to increased escape rate. However, the results of \citet{Baumgardt2002} show this scaling of $\xi$ is small compared to its overall value, so as a first approximation we choose to model the isolated mass loss via equation~(\ref{eq:NdotN}) with constant, best-fitting $\xi_1$ for clusters of any $N_0$.

\subsection{Clusters in a Steady Tidal Field}
\label{s:t3}
 
If a gravitationally bound object is immersed in an external tidal field (i.e. that experienced by a globular cluster orbiting a galaxy), every star within the cluster will experience an acceleration from the galactic potential. However, stars whose orbits take them significantly closer to or further from the central galaxy will experience an increasingly \emph{different} acceleration relative to the mean acceleration over the cluster. Thus, suggested by \citet{VH1957}, there exists a distance whereupon the tidal acceleration due to the galactic potential acting on a star exceeds the acceleration between star and cluster centre. At this distance, a star would become unbound from the cluster, and begin to orbit the parent galaxy independently. \citet{King1962} showed that this limiting distance defines the radius of the Jacobi surface, which is for a point-mass galactic potential and including the centrifugal force given by
\begin{align}
r_{\textrm{J}} = R_{\textrm{G}}\left(\frac{GN\bar{m}}{3M_{\textrm{G}}}\right)^{\frac{1}{3}}. \label{eq:jacobi} 
\end{align}

The dynamical effect of the Jacobi radius will be the addition of an upper limit to the size to which a cluster can expand. By applying our argument that the flux of energy is constant through any surface, the perpetuation of this flux requires stars carrying energy to cross the Jacobi radius, and hence exit the cluster.  The result will be accelerated mass-loss, significantly reducing the lifetime before final dissolution as compared to an isolated cluster. We note also that, on account of the $N^{1/3}$ dependence in equation~(\ref{eq:jacobi}), escaping stars will cause a reduction in the Jacobi radius, which (for self-similar evolution) will cause a shrinking of the entire cluster (i.e. reduction of all Lagrange radii).

In a tidal field we require a new definition of $\xi$ to account for the loss of stars transiting the Jacobi radius. Following the method of \citet{Henon1961}, we assume that the existence of such a boundary will result in a reduced escape velocity, as stars must no longer reach infinity. Using the definition of the escape velocity of a tidally limited cluster ($v_{\rm tidal}$) and the escape velocity from an isolated cluster ($v_{\rm iso}$) from \citet{Spitzer1987}, we find the ratio of escape velocities $\rhrj \propto v^2_{\textrm{tidal}}/v^2_{\textrm{iso}} \propto r/r_{\rm J}$.

From this ratio, \citet{Gieles2008} integrated Maxwellian velocity distributions to find the fraction of stars with sufficient velocity to escape for different ratios $\rhrj$. They showed a good fit was given by $\xi \propto \exp\left(10\rhrj\right)$, although for further analysis we take a simplified form,
\begin{align}
\xi =\frac35\zeta\left(\frac{\rhrj}{\rhrjr}\right)^{z}, \label{eq:xi1}
\end{align}
where $\rhrjr$ defines a reference value of $\rhrj$ and $z$ is a constant power. 
We obtain the factor of $(3/5)\zeta$ present in equation~(\ref{eq:xi1}) as a result of H{\'e}non's interpretation of evolution occurring with constant mean density within clusters. The logarithmic slope relating $\rhrj$ to $N$ is given by,
\begin{align}
\frac{\textrm{d}\ln(\rhrj)}{\textrm{d}\ln N} &=  \frac{\textrm{d}\ln r}{\textrm{d}\ln N}- \frac{\textrm{d}\ln r_{\textrm{J}}}{\textrm{d}\ln N}\label{eq:xi1a} \\
&= \frac53-\frac\zeta\xi \label{eq:xi1aa}
\end{align}
where we have eliminated $\textrm{d} \ln r/ \textrm{d}\ln N$ with equation~(\ref{eq:dNdr}) and derived $\textrm{d}\ln r_{\textrm{J}}/ \textrm{d}\ln N$ from equation~(\ref{eq:jacobi}). From equation~(\ref{eq:xi1aa}) it is evident that for $\xi = (3/5)\zeta$, $\textrm{d}\ln(\rhrj)/\textrm{d}\ln N = 0$ and so $r\propto r_{\textrm{J}}$. It is thus apparent (assuming $\dot{N} < 0$) that if $\xi > (3/5)\zeta$, $r$ will shrink faster than $r_{\textrm{J}}$. Likewise, if $\xi < (3/5)\zeta$, $r$ will shrink slower than $r_{\textrm{J}}$, with the result that $(3/5)\zeta$ is a critical rate whereupon the cluster shrinks with constant density.

We leave the value of $z$ in equation~(\ref{eq:xi1}) to be determined by fitting, although note it's value will effect the variation of $\xi$ with $\rvrj$. The most visible evolutionary effect of the value of $z$ in this regime is to vary the behaviour of $\rvrj$ against $N$, the consequences and effects of which are explained in detail in Appendix~\ref{ap:zx}. 

\subsection{Interpretation of the Jacobi Surface}
\label{s:t4} \label{s:t5}

Thus far, we have considered the escape of individual stars to be a function solely of energy (that is to say, stars with sufficient velocity will always escape). It is however possible that stars with energy slightly in excess of the critical energy remain bound, owing to geometric constraints. Specifically, escape is only possible through `apertures' around the $L_1$ and $L_2$ Lagrange points, over which the gradient of the potential is sufficiently shallow for stars with energy only slightly greater than critical to escape. It follows that stars with greater energy can escape over a larger region, whilst those with less energy will experience a smaller escape aperture.

As a result of these apertures, \citet{Fukushige2000} characterised the relationship between $t_{\textrm{esc}}$ (the time taken for any given star to escape) and the excess energy $(E - E_{\textrm{crit}})$ (where $E_{\textrm{crit}}$ is the exact energy required for escape) to be $t_{\textrm{esc}} \propto \left[E_{\textrm{crit}}/(E-E_{\textrm{crit}})\right]^2$, from which \citet{Baumgardt2001} obtained a form for the lifetime of a cluster, 
\begin{align}
t_{\textrm{ev}} &\propto t_{\textrm{rh}}^{x}t_{\textrm{esc}}^{1-x}, \\
&\propto t_{\textrm{rh}}\left(\frac{N\ln(\gamma N_1)}{N_1\ln(\gamma N)}\right)^{x-1}, \label{eq:tesc}
\end{align} 
where $N_1$ is a separate scaling value of $N$ that shall be determined from comparisons to $N$-body simulations. It follows that, since a star with {$E \gtrsim E_{\textrm{crit}}$} will experience a delay before finally exceeding the Jacobi radius, we should account for these stragglers in our evaporation rate. We therefore combine the escape timescale for stars (equation~\ref{eq:tesc}), with the rate through which evaporation causes stars to evaporate from the cluster (equation~\ref{eq:xi1}), and hence obtain
\begin{align}
\xi = \frac35\zeta P(\rvrj,N), \label{eq:xi2}
\end{align}
where
\begin{align}
P(\rvrj,N) = \left(\frac{\rvrj}{\rvrjr}\right)^{z}\left(\frac{N\ln(\gamma N_1)}{N_1\ln(\gamma N)}\right)^{1-x}, \label{eq:PP}
\end{align}
an improved form for the dimensionless evaporation rate of a cluster in a tidal field.

\begin{figure*}
 \centering
   \includegraphics[width=170mm]{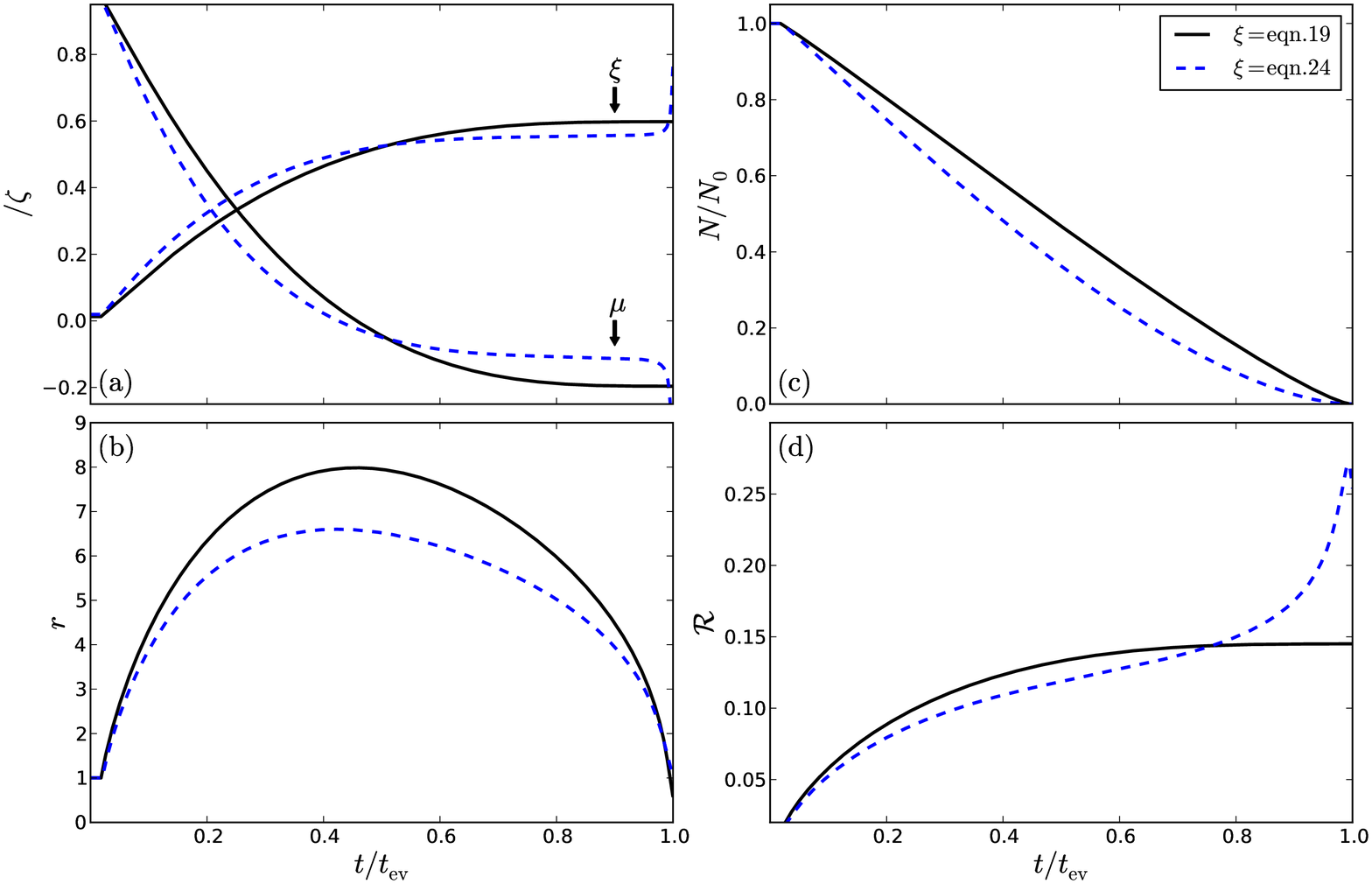}
 \caption{Predicted evolution of star cluster parameters according to the models outlined in section~\ref{s:t3}, and computed using the technique outlined in section~\ref{s:3}. (a) shows the evolution of parameters $\xi$ and $\mu$ in time, scaled to $\zeta$, while (b), (c), and (d) show the number of remaining stars, half-mass radius, and ratio of $\rvrj$ respectively. Early divergence between the models is principally an effect of the scaling with $N^{1-x}$, whist the late divergence between the models (especially noticeable in panel (d)) is an effect of the Coulomb logarithm, which becomes dominant at low $N$. For the illustrations, $\zeta = 0.1$, $N_1 = 10000$, $\rvrjr =0.145$, $\gamma = 0.11$ and $x = 0.75$, for $N=65536$ clusters with $\rvrjo = 1/100$. For such clusters, the lifetime $t_{\rm ev} \approx 700000$ $ N$-body times, with core collapse occurring after $\sim12000$ $ N$-body times. If these clusters were to consist of $0.5M_{\odot}$ stars, an initial $r$ of $1{\rm pc}$ would imply an expected lifetime of $\sim 60$Gyr, with core collapse occurring at $1$Gyr.}
 \label{f:tilim}
\end{figure*}

\section{A full model for the life cycle}
\label{s:4}
Sections~\ref{s:t2} and~\ref{s:t5} established physical arguments to predict evolution of the mass and radius of clusters undergoing either expansion or contraction. It is apparent that a realistic cluster starting it's evolution with $r \ll r_{\textrm{J}}$ will expand to fill it's Jacobi radius, eventually becoming tidally limited and thus experiencing both regimes. We therefore attempt to merge these two formulae into a single unified model, encompassing the entire lifetime of star clusters.

For clarity, we shall refer to the tidally limited escape rate (equation~\ref{eq:xi2}) as $\xi_{\textrm{tidal}}$. We also assume that both mechanisms for mass loss are viable channels through which stars can escape over the entire lifespan of the cluster. However, the extent to which the two mechanisms are significant will vary as the cluster expands, since early times are likely to be dominated by tidal field independent mass-loss mechanisms and late times by tidal mechanisms. To account for this variation we take the factor $P(\rvrj,N)$ from $\xi_{\textrm{tidal}}$ (used to represent the $\rvrj$ and $N$ dependencies in this term), and which varies such that $0<P(\rvrj,N)<1$. We now assume that the variation of $\xi_1$ (the mass loss of isolated clusters) will be (approximately) opposite to that of $\xi_{\textrm{tidal}}$, and thus let $\left(1 - P(\rvrj,N)\right)$ represent the $\rvrj$ and $N$ dependence in $\xi_1$. We therefore write,
\begin{align}
\xi &= \xi_1\left(1 - P(\rvrj,N)\right)+\xi_{\textrm{tidal}} \\
 &= \xi_1\left(1 - P(\rvrj,N)\right)+\frac35\zeta P(\rvrj,N). \label{eq:xi3}
\end{align}
Although equation~(\ref{eq:xi3}) represents an essentially simple interpretation of the major mass-loss mechanisms present in star clusters, we believe an interpretation of this nature is a sufficient description if the two mechanisms are both present throughout the lifetime. 

Using equation~(\ref{eq:xi3}), we now outline the operation of our numerical integration code \textsc{EMACSS}. The principle properties, $N$ and $r$, are recovered by application of a 4th order Runge-Kutta numerical integration kernel to our defining equations~(\ref{eq:NdotN}) and~(\ref{eq:rdotr}) with $\xi$ and $\mu$ calculated appropriately at each integration step, and constant $\zeta$. The duration of each time step is set to be $0.1t_{\textrm{rh}}$, which we find to be a reasonable compromise between speed and accuracy in the model. This fraction is appropriate as $\xi$ and $\mu$ are defined as instantaneous values for the rate of change of $N$ and $r$ per half-mass relaxation time. Thus $t_{\textrm{rh}}$ forms an upper limit to time step, while an overly short time step is liable to be affected by numerical inaccuracies.

\begin{figure*}
 \centering 
   \includegraphics[width=170mm]{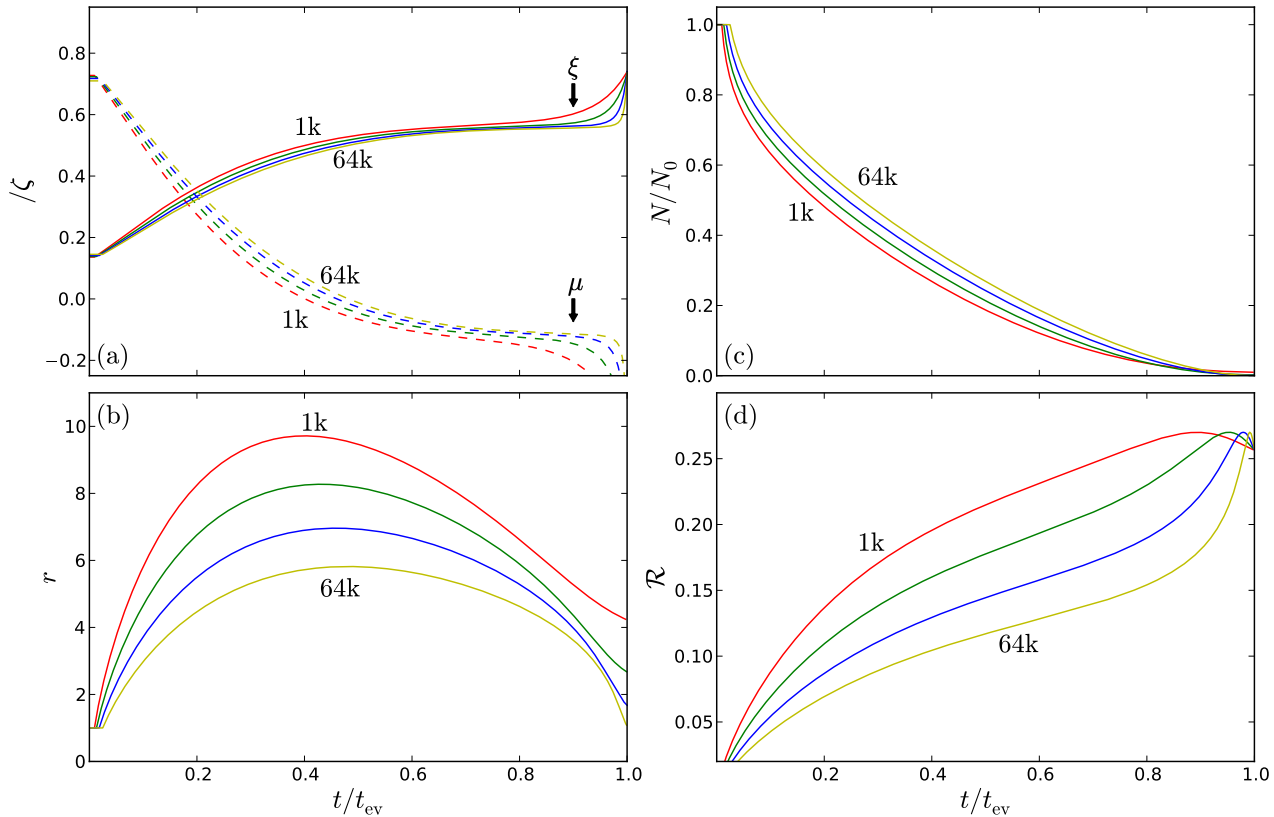}
 \caption{The predicted evolution of star-clusters containing varying $N_0$, using equation~(\ref{eq:xi3}). (a) shows the evolution of parameters $\xi$ (solid) and $\mu$ (dashed) in time, scaled to $\zeta$, while (b), (c), and (d) show the number of remaining stars, half-mass radius, and ratio $\rvrj$ respectively. The multiple lines denote clusters with $N_0$ determined by successive factors of four, between 1024 (labeled 1k) and 65536 (labeled 64k). The two post-collapse regimes are most visible in (b), with early mass loss on account of $\xi_1$ and late due to $\xi_{\textrm{tidal}}$. A maximum size is achieved at $t/t_{\textrm{ev}} \approx 0.45$ (dependent on initial $\rvrj$), whereupon $\mu$ becomes negative and the cluster begins to shrink on account of the decrease in Jacobi radius due to decreasing cluster mass. The final changes in the cluster (after around $t/t_{\textrm{ev}} = 0.9$) are likely accounted for by the model breaking down, as here $t_{\textrm{cr}} \approx t_{\textrm{rh}}$, and thus $\zeta$ non-constant. The similarity of evolution in panel (b) is likely an effect of weakness of the scaling of $\xi$ with $N$. These illustrations use the same parameters as figures~\ref{f:notidecomp} and~\ref{f:tilim}.}
 \label{f:fullfig}
\end{figure*}

At each integration time $t_i$, our procedure is as follows:
\begin{enumerate}
\item{Characteristic properties of the cluster ($r_{\textrm{J}}$ and $t_{\textrm{rh}}$, required for the dynamical evolution), are evaluated from $N(t_i)$ and $r(t_i)$.}
\item{$\xi$ (equation~\ref{eq:xi3}) is evaluated for instantaneous values of $N(t_i)$ and $r(t_i)$, and used to  calculate $\mu$ via equation~(\ref{eq:relate}).}
\item{A Runge-Kutta integration step is applied to equations~(\ref{eq:NdotN}) and~(\ref{eq:rdotr}) using parameters derived in stages (i) and (ii), with these properties re-evaluated as appropriate. In this stage $N(t_{i+1})$ and $r(t_{i+1})$ are recovered.}
\item{$N(t_{i+1})$, $r(t_{i+1})$, and other properties (as required) are output.}
\item{Steps (i) through (iv) are repeated until $N \le 200$, at which point the half-mass crossing time $t_{\textrm{cr}} \approx t_{\textrm{rh}}$, and balanced evolution is no longer a valid assumption.}
\end{enumerate}
 For simplicity, we introduce $N$-body units (such that $G = N\bar{m} = -4E = r = 1$; \citealt{Heggie1986}), although conversion to physical units is trivial (\textsc{EMACSS}, by default, outputs both). We use the above procedure to illustrate sections~\ref{s:t2} and~\ref{s:t4}. In figure~\ref{f:tilim}, we demonstrate the evolution of a cluster with $\xi$ defined by equations~(\ref{eq:xi1})and~(\ref{eq:xi2}). Meanwhile, figure~\ref{f:fullfig} demonstrates the evolution of the system using a $\xi$ defined by equation~(\ref{eq:xi3}), for a variety of $N_0$.

The form of equation~(\ref{eq:xi3}), and hence the model \textsc{EMACSS} are dependent upon several free parameters (see table~\ref{t:allparam}). However, degeneracy amongst these parameters is such that we can adopt an appropriate value for $\rvrjr$, and interpret the scaling factor $N_1$ to define an `ideal' clusters for which this value is exactly correct. We therefore choose $\rvrjr = 0.145$ (\citealt{Henon1965-1}). We additionally note that $x$, $z$ and $N_1$ all demonstrate a high degree of covariance (see Appendix~\ref{ap:zx}), although are not totally degenerate. However, to the level of accuracy for which this model is intended, this covariance is sufficient that we can eliminate a further parameter without loss of generality. To this end, we choose $x = 0.75$ \citet{Baumgardt2001}, and choose a value of $\gamma = 0.11$ from the results of \citet{Giersz1994} for the Coulomb argument, leaving our model dependent only upon  $\xi_1$, $N_1$, $z$ and $\zeta$. These remaining (free) parameters are determined by calibration of our model against a database of $N$-body simulations with differing initial conditions, which we describe below.

\begin{table}
  \centering
  \caption{Definitions of the parameters used to describe cluster evolution in \textsc{EMACSS}.}
  \begin{tabular}{c p{50mm}}
    \hline
    Variable & Definition \\
    \hline
    $\zeta$  & Fractional change in energy per (half-mass) relaxation time.\\
    $\xi$  & Dimensionless escape rate.\\
    $\mu$  & Dimensionless expansion rate.\\
    $\rhrj$  & Ratio of radius to Jacobi radius. \\
    $t_0$ & Time taken for balanced evolution to begin in isolated clusters. \\
    \\
    \hline
    Parameter & Definition \\
    \hline
    $\xi_1$  & Dimensionless escape rate of an isolated cluster.\\
    $\rhrjr$  & Ratio of radius to Jacobi radius for \citet{Henon1965-1} models ($\rhrjr = 0.145$). \\
    $N_1$ & `Ideal cluster' for which $\rhrjr$ is exactly correct.\\
    $z$ & Scaling of $\xi$ with $N$ around $\rhrjr$.\\
    $x$  & Variation of $\xi$ with $t_{\rm rh}$.\\
    $\gamma$  & Argument of the coulomb logarithm. \\
    $t_{\rm cc}$  & Time in which core collapse is fully completed ($> t_0$). \\
    \hline
  \end{tabular}
  \label{t:allparam} 
\end{table}

\section{Description of the $\textbf{N}$-body Simulations} 
\label{s:3}

We produced a number of $N$-body simulations using the collisional fourth order Hermite $N$-body code \textsc{Nbody6} (\citealt{Makino1992,Aarseth1999,Aarseth2003}), on Intel Core i7 computers. Parallelisation and GPU acceleration was provided by NVDIA GTX GeForce graphics processing units. Our simulations evolved clusters containing between $N_0=1024$ and $N_0=65536$ equal-mass stars, with a separate series of simulations carried out at every intervening factor of two.  The clusters were initially described by \citet{Plummer1911} models, with initial $r_{\textrm{h}} = 0.78$, and henceforth allowed to evolve until $N \le 200$ (approximately the time of final dissolution, $t_{\textrm{ev}}$) .

We defined the ambient tidal field for each simulation in terms of initial ratio of half-mass to Jacobi radius $\rhrjo$, using initial ratios $1/30$ and $1/100$. The tidal field implemented was that of a point-mass galaxy, with appropriate $M_{\textrm{G}}$ and $R_{\textrm{G}}$ to give the required $r_{\textrm{J}}$ via equation~(\ref{eq:jacobi}). The numbers of our simulations are summarised into table~\ref{t:runs}, along with several simulations of isolated clusters kindly provided by \citet{Baumgardt2002}.

\begin{table}
  \centering
    \caption{The number of simulations carried out for each set of conditions $N$, $\rhrjo$. The isolated simulations were provided by \citet{Baumgardt2002}.}
  \begin{tabular}{cccc}
  \hline
    & \multicolumn{3}{c}{$\rhrjo$} \\
    $N$ & $1/30$ & $1/100$ & Isolated \\
    \hline
    1024 & 64 & 64 & 3\\ 
    2048 &  32 & 32 & 1\\ 
    4096 & 16 & 16 & 1\\
    8192 & 8 & 8 & 1\\
    16384 & 4 & 4 & -\\
    32768 & 2 & 2 & -\\
    65336 & 1 & 1 & - \\
    \hline 
  \end{tabular}
  \label{t:runs} 
\end{table}

For isolated clusters, unbound stars were defined as those with energy in excess of that required for escape (i.e. positive energy). These escaping stars remained within the simulation until they reached $20r$, in order to retain their gravitational influence upon bound stars. Meanwhile, in tidally limited clusters, the Jacobi radius was calculated iteratively using equation~(\ref{eq:jacobi}), and unbound stars were defined as those outside $r_{\textrm{J}}$. These stars were removed from the simulation upon exceeding $2r_{\textrm{J}}$. This criterion for escape will consider stars with $E>E_{\textrm{crit}}$ but $r<r_{\textrm{J}}$ to be bound\footnote{An alternative, energy based formalism has been presented by \citet{Lee1987}, although since we seek to include the effect of escaping stars into our model, we choose instead to apply a radial criteria for escape.}. In both situations, a list of (bound) particles ordered by radial distance was used to determine the half-mass radius.

We finally measured the energy budget of bound stars at each time-step. For this purpose, we computed the `external' energy (kinetic and potential components of single stars and the centres of mass for multiples, \citealt{Giersz1997}) separately from the `internal' energy of particles (that stored in binaries and multiples). In addition, we recorded the cumulative energy loss from the cluster for single star escapers and escaping multiple systems. Using these data, we were able to track the change of energy, and hence (given our premise of balanced evolution) the energy flux through any given radius.

\section{Results}
\label{s:res}
The procedure for the calibration and verification of our prescription against $N$-body data is divided into three principle sections.

As a first stage, we begin by examining the flow of energy throughout the cluster. To this end, in section~\ref{s:m1} we use the energy budget of bound stars to determine $\dot{E}/E$ by numerical differentiation of $E$ throughout the evolution. We then calculate $t_{\textrm{rh}}$ through equation~(\ref{eq:relax}), and hence determine $\zeta$ through rearranging equation~(\ref{eq:EdotE1}). This quantity is the `driver' defining the evolution of a cluster, and hence forms a significant limit on the speed through which a cluster evolves and the extent of the total lifetime.

Section~\ref{s:m2} contains the results of our calibration of \textsc{EMACSS} against $N$-body simulations of isolated clusters. In these cases $\xi = \xi_1$ since $r_{\textrm{J}}$ is infinite and $\xi_{\textrm{tidal}}=0$. Accordingly, we recover a value for $\xi_1$ that best represents the evolution of isolated clusters. We then apply \textsc{EMACSS} to our $N$-body simulations of tidally limited clusters, in section~\ref{s:m3}, and thus assign values for the remaining parameters, $N_1$ and $z$. Using these values, our prescription is able to concisely express the evolution of $N$ and $r$ for clusters throughout their entire life cycle. 

In each case, our best-fitting parameters are recovered by a custom designed Markov Chain Monte Carlo (\citealt{Metropolis1953}) routine. For simplicity, we take a series of sample data encompassing the entire evolution of our $N$-body simulations, and apply a Gaussian likelihood function to compare the samples' $N$ and $r$ values to the equivalent $N$ and $r$ values predicted by \textsc{EMACSS}. Uncertainties throughout are calculated from the standard deviation of our $N$-body simulations, but limited to a minimum of 10\% of the absolute value (the level of accuracy to which we believe our model to be valid). By marginalising the parameters we obtain best fitting parameters for each quantity in the model, and can estimate the uncertainty of each.

\subsection{The flow of energy}
\label{s:m1}

\begin{figure}
\centering
\includegraphics[width=80mm]{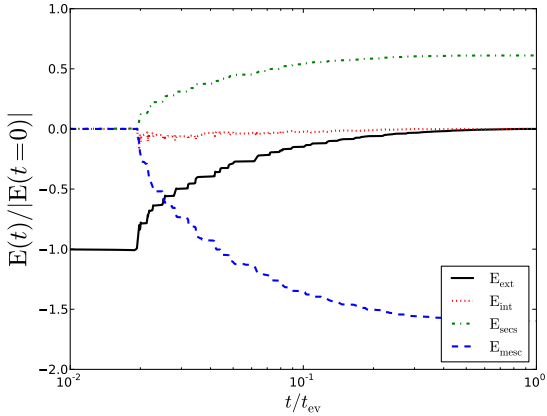}
\caption{Energy budget for a representative $N=65536$ run evolving in a tidal field with initial $r/r_{\textrm{J}} = 1/100$. The black (solid) line corresponds to the external energy of stars in the cluster, the red (dotted) line is the internal energy, the green (dot-dash) line the cumulative energy of singular escapers and the blue (dashed) line the cumulative energy of escaped multiples. The early (invariant) relationship between energy and time is the pre-core collapse evolution, and thus before the start of the energy evolution of the cluster. The downward `spikes' in the internal energy correspond to three body interactions, in which a single star is ejected with positive (kinetic) energy. Meanwhile, the binary involved in such interactions hardens, before finally becoming sufficiently hard to escape, carrying away it's (negative) binding energy. Effects of this type can be seen to occur intermittently throughout the life cycle.}
\label{f:escaping}
\end{figure}  

Figure~\ref{f:escaping} shows the evolution of the three principle components of the energy budget typical for our $N$-body simulations. The quantities expressed are the total external energy $E_{\textrm{ext}}$ of bound stars within the system (kinetic and  potential components), the total internal energy  $E_{\textrm{int}}$ (that in binaries and multiples) and the cumulative total energy carried away by singular escaping stars ($E_{\textrm{sesc}}$) or multiples ($E_{\textrm{mesc}}$). As the cluster is formed in a nearly isolated state, early escapers have positive  kinetic energy, since they are ejected via close encounters with binary or higher order multiples. Accordingly, corresponding to each escaping (single) star is a decrease in the internal energy of the cluster, as the multiple system involved in such encounters will become more tightly bound. Through interactions, a multiple will eventually be in turn ejected from the cluster, and hence carry away it's negative (binding) energy. 

Stars ejected late in the life cycle are typically less energetic than those ejected earlier. This would arise on two accounts; firstly, as demonstrated in figure~\ref{f:escaping}, much energy is lost during expansion, and hence the stars of a tidally limited cluster retain less energy. Secondly, the escape velocity of a Roche lobe filling cluster is reduced by presence of a Jacobi radius, with the result that outlying stars with negative energy can exit the cluster. 

We begin our analysis by finding the value of $\zeta$ (see section~\ref{s:EvSC}), as this is the defining characteristics upon which our model is built. For this purpose, we use the variation of the external energy $E_{\textrm{ext}}$ as a function of time, and sequentially bin these data logarithmically in energy. In each bin, we approximate the change in the log energy over the log time extent of the bin to be linear, and hence determine the gradient ${\rm d}\log (-E)/{\rm d} \log t$ within each bin. 

We then estimate the mean $\bar{t}_{\textrm{rh}}$ corresponding to each bin via equation~(\ref{eq:relax}).We have previously chosen to work in terms of $N$ and virial radius $r$, although note $t_{\rm rh}$ is defined in terms of $r_{\rm h}$. We therefore first make the assumption that $r_{\rm h} = r$ to determine a value for $t_{\rm rh}$, before rearranging equation~(\ref{eq:EdotE1}) such that
\begin{align}
\zeta_{i} = \frac{\bar{t}_{{\rm rh,}i}}{\bar{t}_{i}}\left|\frac{{\rm d}\log (-E_{i})}{{\rm d}\log t_i}\right| \label{eq:rearrh}
\end{align}
for an arbitrary bin $i$. We finally substitute $\bar{t}_{\textrm{rh}}$, $\bar{t}_i$, and ${\rm d}\log (-E_i)/{\rm d} \log t_i$ into equation~(\ref{eq:rearrh}) to obtain a distinct value of $\zeta$ corresponding to each bin. For completeness, we also measure the value of $\zeta$ for a relaxation time recovered using the measured $r_{\rm h}$ of $N$-body simulations, and compare the resultant two values of $\zeta$.

\begin{figure*} 
\centering
   \includegraphics[width=170mm]{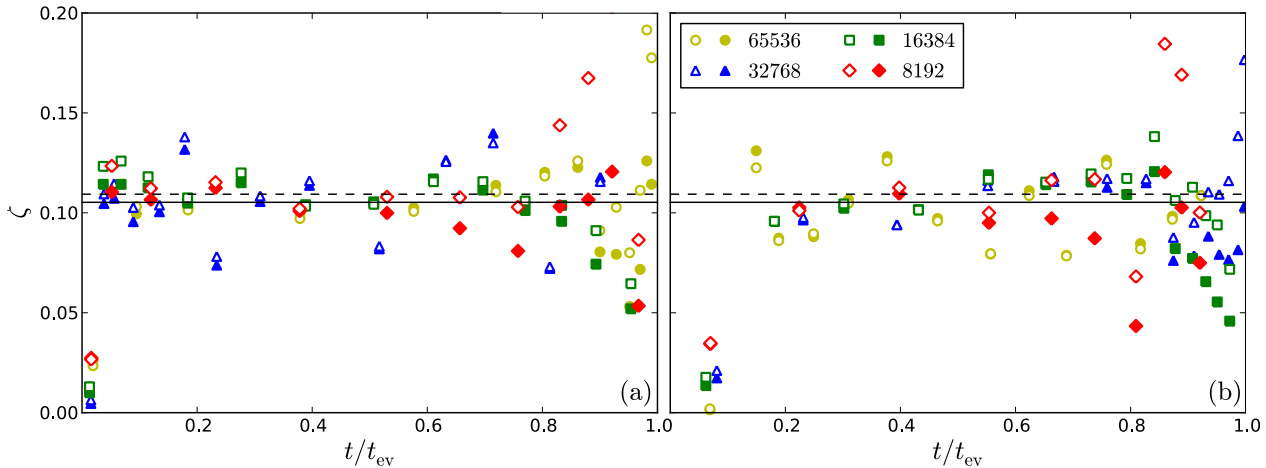}
\caption{Variation in $\zeta$ as a function of time for a variety of runs with different $N_0$. Only runs with $N_0 \ge 8192$ stars are plotted, as the stochastic nature of the energy makes the scatter of smaller runs progressively worse. Filled shapes and the dashed line correspond to measurements of $\zeta$ where we have assumed $r_{\rm h} = r$, while unfilled shapes (and the solid line) corresponds to measurements of $\zeta$ where $r_{\rm h}$ is measured directly from $N$-body simulations. Panel (a) shows the runs carried out with an initial $\rhrj=1/100$ while panel (b) has initial $\rhrj=1/30$. The early (systematically low) points are likely caused by no energy loss occurring during core collapse. There is a possible downward slope at late times, which may be a result of the break down of balanced evolution.}
\label{f:zeta}
\end{figure*}

Figure~\ref{f:zeta} shows the variation in $\zeta$ with time, for each of our initial tidal field strengths. Although in each case numerical differentiation has lead to significant scatter, we find that both measurements of $\zeta$ vary around constant values, $\zeta \approx 0.111$ if $t_{\rm rh}$ is measured from $r$ and $\zeta \approx 0.105$ if $t_{\rm rh}$ is measured directly from $r_{\rm h}$. These results are similar to previously calculated values for H{\'e}non's models ($\zeta = 0.1$ (\citealt{Goodman1989}), $\zeta=0.14$ (\citealt{Gao1991}), $\zeta = 0.0926$ for isolated clusters, $\zeta = 0.0743$ for tidally limited clusters (\citealt{Gieles2011})). By contrast, there is some indication that the smooth flow does not occur during two stages of evolution; the first of these corresponds our earliest (furthest left) data, which are significantly lower than the average flow. This is the result of pre-collapse evolution, during which energy does not change and hence $\zeta=0$. The second (toward the end of evolution) corresponds to the final break down of the smooth energy flow, as at these times $t_{\textrm{rh}} \approx t_{\textrm{cr}}$ and the redistribution of energy becomes increasingly random throughout star clusters.

\subsection{Evolution of isolated clusters}
\label{s:m2}
We now begin our investigation of \textsc{EMACSS}'s ability to reproduce the evolution of star clusters, beginning with the simple case of a cluster evolving in isolation. In the previous section we demonstrated $\zeta  = 0.105$\footnote{Using $\zeta$ measured from $N$-body simulations.}, which we now adopt as the energy flow driving the evolution of $N$ and $r$. We use the numerical integrator described in section~\ref{s:4} to evolve our model cluster, defining the galaxy around which our simulated cluster is `orbiting' to have no mass, and hence our cluster to be isolated from tidal effects. The resultant evolutionary tracks of $N$ and $r$ are compared against those of $N$-body simulations of isolated clusters.

For calibrating our prescription we use $\xi_1$ as a free parameter, and introduce two further parameters, $f_N$ and $f_r$, the fractional change in $N$ and $r$ during core-collapse. The time taken for this collapse (assuming an initial \citet{Plummer1911} sphere) has been reported to be 17.6 initial relaxation times ($t_{\textrm{rh,0}}$) (\citealt{Takahashi1995,Drukier1999}), in which we expect relatively small changes in $N$ and $r$. However, in order that the entire collapse (and core bounce) is complete when we begin to apply our prescription, we choose to use a time offset of $20t_{\textrm{rh,0}}$, and define $f_N$ and $f_r$ to be the fractional changes that have occurred within the cluster prior to this time. Naturally, it is therefore also possible to project our prescription backward until the time of core collapse itself, and hence model the entire post-collapse evolution of a star cluster. We find values for $f_N$ and $f_r$ from linear interpolation of Baumgardt's $N$-body data at $20t_{\textrm{rh,0}}$, and calibrate our prescription using the Monte Carlo algorithm described in section~\ref{s:4}. 

\begin{figure*} 
\centering
    \includegraphics[width=170mm]{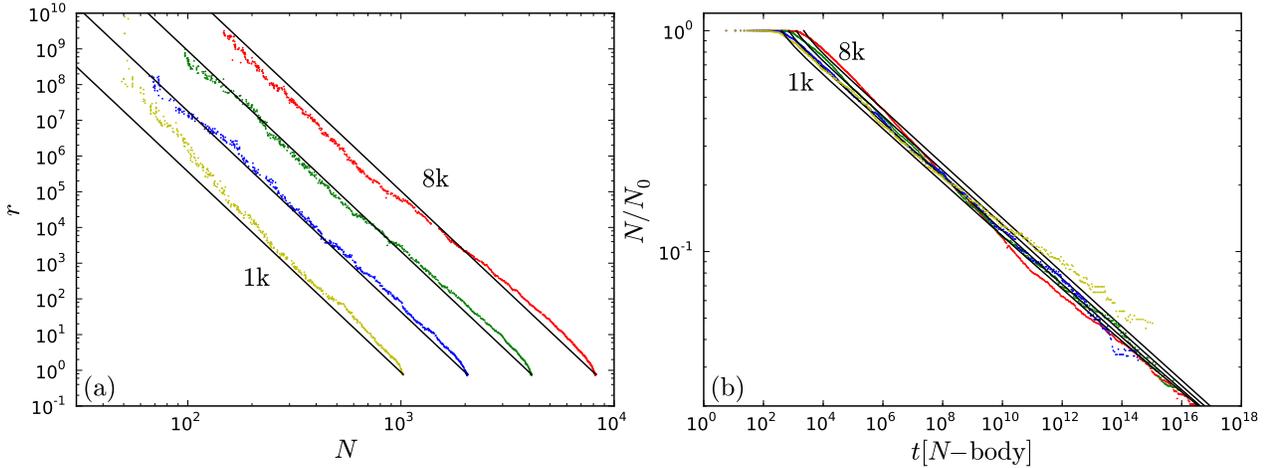}
\caption{(a) The relationship between $N$ and $r$ for clusters of different $N_0$. (b) The evolution of $N$ (scaled to the initial cluster size $N_0$). The scattered points denote the evolution of clusters with $N_0$ determined by successive factors of two, between 1024 (labeled 1k) and 8192 (labeled 8k), while the black lines are produced using our prescription with our best fitting parameters. Our (best-fitting) prescription provides a general indication of the rate at which an isolated cluster expands and evaporates, although does not reproduce the crossing point found by \citet{Baumgardt2002}. There is some evidence (explained below) that the systematic discrepancies between our prescription and the $N$-body data are effects of isolated clusters not being fully self-similar throughout the entirety of their evolution.} 
\label{f:notides} 
\end{figure*} 

Figure~\ref{f:notides} shows the evolutionary tracks predicted by our best-fitting parameters for the entire lifetime of the cluster. Our best fitting models occur where $\xi_1 = 0.0141$, consistent previous findings ($\xi_1 \approx 0.01$, \citealt{Heggie1992-2,Baumgardt2002}). Our interpolation demonstrates a mean ($N$ independent) expansion during core-collapse $f_r = 1.81$, with the majority of this expansion occurring immediately after the predicted $17.6t_{\rm rh,0}$ core collapse time. Meanwhile, mass-loss occurs such that  $f_N = 0.95$. From figure~\ref{f:notides}, it is apparent that this mass-loss begins gradually, building up over a period of time to approximately constant $\xi_1$ observed for the majority of the life cycle. It is clear therefore that backward projection of our prescription will result in an instantaneous start to mass-loss, an effect that we do not believe to be  physically likely. We therefore conclude that the early evolution -- directly after core collapse -- will be poorly described by this prescription

The principle discrepancies between our best fitting prescription and comparison $N$-body data consist of some mild variation of log-gradient $\textrm{d}\log r/\textrm{d}\log N$; while we represent the evaporation rate as being constant, the $N$-body data displays a more complex evolution, a point noted by \citet{Baumgardt2002}. In this paper the authors demonstrate that expansion is not strictly self-similar for an isolated cluster, although they find similar scaling between expansion and mass-loss at an inner Lagrange radius $(10\%-20\%)$. We consider this to be a consequence of an increased rate of ejection causing encounters on account of increased core density for larger $N$ systems, and hence interpret the discrepancies in our gradient to be a result of clusters not being self-similar throughout their evolution, although note our current self-similar description does provide a good first order interpretation.

We demonstrate the veracity of our best fitting parameters by considering our first approximation of constant Coulomb logarithm. Here, $N$ should go as a negative power-law with time (equation~\ref{eq:Nv1}), with the power $\nu$ defined by equation~(\ref{eq:nu}). Using $\zeta  = 0.105$ and $\xi_1 = 0.0141$, we compare the initial mass-loss rate from our results to the mass-loss rates found by \citet{Baumgardt2002} for our range of $N_0$. Our measurements give that $\nu = 0.130$, independent of $N_0$. This value of $\nu$ demonstrates excellent agreement with Baumgardt's values of $\nu$ calculated by the evolution of $N$ ($0.13 \le \nu \le 0.14$ for $1024 \le N \le 8192$), although less good agreement with Baumgardt's $\nu$ calculated by the evolution of $r$ ($0.02 \le \nu \le 0.09$ for $1024 \le N \le 8192$). We conclude therefore that equation~(\ref{eq:Nv1}) can be used to calculate the relationship between $N$ and time, but that this scaling cannot simultaneously reproduce the evolution of $r$ with comparable precision, as $N$ and $r$ do not scale with the same power-law.

We finally examine our definition of $t_0$ (equation~\ref{eq:t0}), which we have described as being the time required for a model of non-zero $r_0$ to join the track predicted by balanced evolution (where $t=r=0$). Using the value of $\xi$ and $\zeta$ above, we find $t_0 \approx 9.2t_{\textrm{rh,0}}$, a time shorter than the core-collapse of a Plummer profile. This value does demonstrate the dependence of the length of the core-collapse upon initial conditions, but does not give any insight into the subsequent global evolution.

\subsection{Evolution in a tidal field} 
\label{s:m3}
A cluster in a tidal field will first experience an expansion phase, before becoming significantly affected by the tidal field and entering a contraction phase. We assume the expansion phase to be similar to that which we have previously demonstrated for an isolated cluster, and hence expect a value of $\xi_1 \simeq 0.014$. We initially set $\zeta = 0.111$ (using our assumption that $r_{\rm h} = r$ when calculating the relaxation time) and $t_{\textrm{cc}} = 20.0t_{\textrm{rh,0}}$.

\begin{figure}
\centering
    \includegraphics[width=85mm]{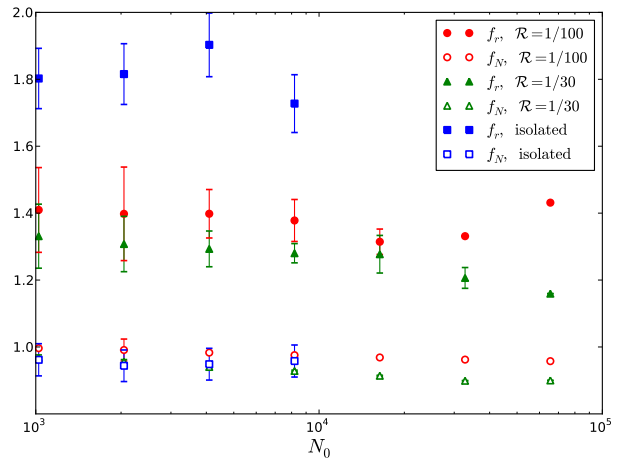}
\caption{Mass-loss ($f_N$) and expansion ($f_ r$) of clusters pre core-collapse. The most significant dependency is that with $\rhrjo$ (on account of the overall escape energy). There is a small $N_0$ dependency visible in $f_N$, although no particular relationship between $f_r$ and $N_0$.} 
\label{f:pre-collapse} 
\end{figure}

For tidally limited clusters, we expect the mass-loss prior to core collapse to be more significant than that found for isolated clusters. From the evidence of equation~(\ref{eq:xi2}), we expect this mass-loss to be a function of $\rvrj$, and therefore allow a unique $f_N$ and $f_r$ for each simulation. Accordingly, we measure these values of $f_N$ and $f_r$ by linear interpolation of our $N$-body data at $20t_{\rm rh,0}$, and consequently recover the specific fractional changes in each series of simulated clusters during core-collapse. These fractional changes in the size and mass of clusters are demonstrated in figure~\ref{f:pre-collapse}.

The extent of mass-loss and expansion during core-collapse shown in figure~\ref{f:pre-collapse} reveals a marginally greater expansion but lower mass-loss for higher $\rhrjo$. In each case, simulated clusters in stronger initial tidal fields expand less, but loose more mass than those in weaker tidal fields. Additionally, figure~\ref{f:pre-collapse} suggests $f_N$ is only weakly dependant on $N_0$, implying that the fraction of stars ejected prior to core collapse is not significantly dependant upon the total number. We use these direct measurements of expansion and mass-loss for the purposes of our fitting.

We employ our Monte Carlo algorithm to simultaneously fit our prescription to each of our tidally limited $N$-body simulations (table~\ref{t:runs}), comparing the evolution of $\xi_1$, $N$ and $r$. The resultant joint posterior probability density demonstrates a narrow peak (as shown in Appendix~\ref{ap:pos}). This is most probably a result of covariance within our data, since each datum will be highly correlated to those around it. We therefore take our repeated database of $N=4096$ and $N=8192$ simulations, and independently fit each of these with our prescription, producing a unique value of $N_1$ and $z$. We find the standard deviation of these individual $N_1$ fits to be about $40\%$ of the mean value, while the standard deviation of the $z$ fits is about $15\%$. We hence interpret these standard deviations as estimates of the uncertainty on these parameters.

\begin{figure*}
\centering
    \includegraphics[width=170mm]{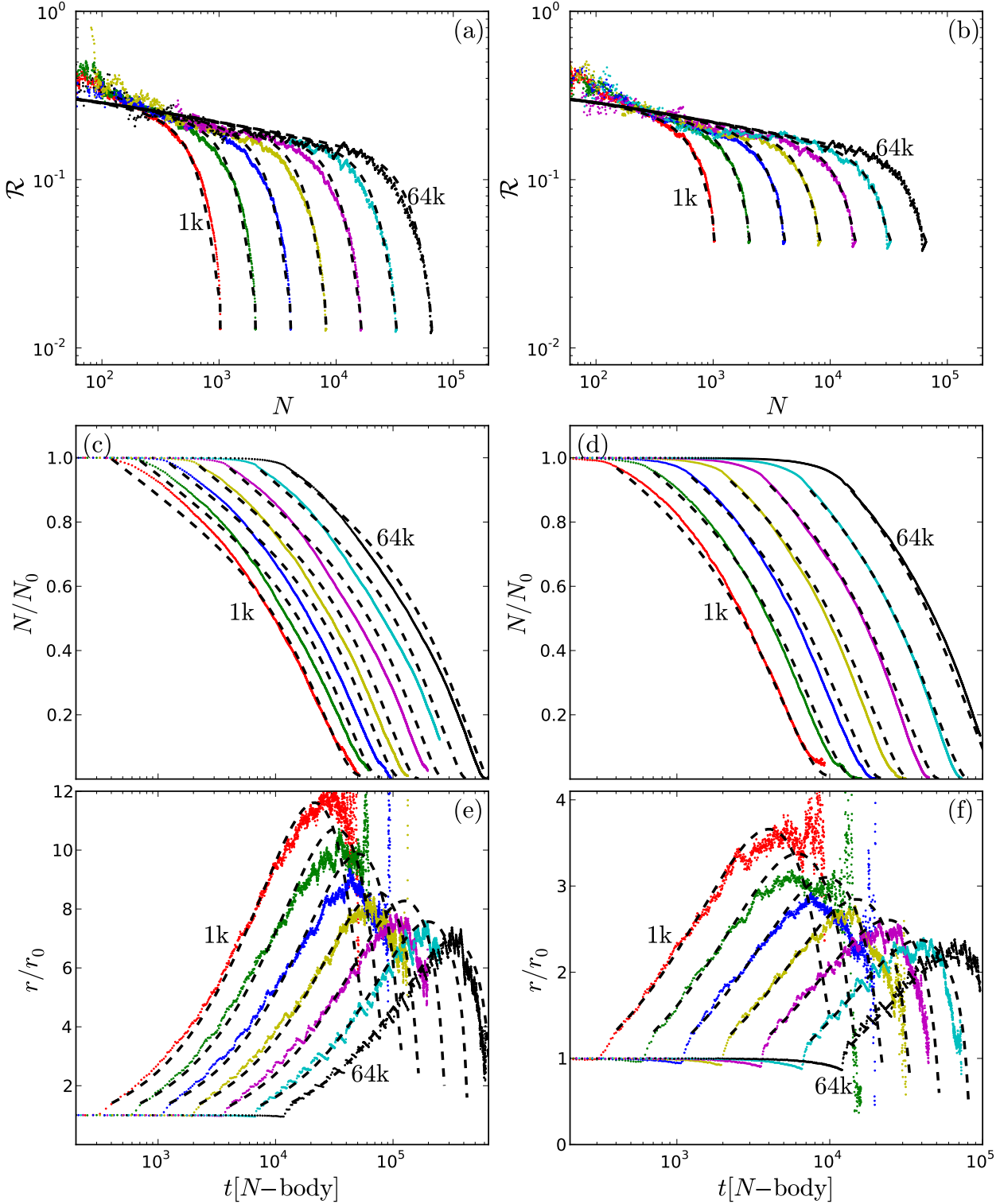} 
\caption{The evolution of star clusters in point mass tidal fields. The scattered points denote the evolution of clusters with $N_0$ determined by successive factors of two, between 1024 (labeled 1k) and 65536 (labeled 64k), while the black (dotted) lines are produced using our prescription with our best fitting parameters. The left hand column corresponds to an initial $\rhrjo$ of $1/100$, while for the right hand $\rhrjo = 1/30$. (a,b) $N$ against $\rvrj$. The movement in the $N-\rvrj$ plane is consistent for all $N$-body simulations; the tracks converge once the cluster is tidally limited, with expansion dominant until this transition. (c,d) $N$ as a function of time. (e,f) Radius $r$ as a function of time. The expansion and contraction stages are accurately reproduced, although the fit for the model is less good during the transition between these regimes (i.e. peak $r$). The oscillations observed in the later evolution of $r$ are possibly indications of gravothermal oscillations or stochastic noise in the core. The model once again fails for low $N$. For $0.5{\rm M}_{\odot}$ stars, an initial $r$ of $1$pc would give an expected lifetime of 35Gyr for the $N=1024,\; \rhrjo = 1/100$ simulations (core collapse after $260$Myr), and $55$Gyr for the $N=65536,\; \rhrjo = 1/100$ simulation (core collapse after $1$Gyr). Meanwhile, the expected lifetime of the  $N=1024,\; \rhrjo = 1/30$ simulations is around $6$Gyr, and that of the $N=65536,\; \rhrjo = 1/30$ simulation is around $11$Gyr.}
\label{f:fulltides} 
\end{figure*}

The best fitting parameters we recover are shown in table~\ref{t:comp}, whilst we over-plot the predicted evolution from our prescription to $N$-body data in figure~\ref{f:fulltides} (dotted lines). For the sake of completeness, we additionally show example joint and marginal posteriors distributions recovered by our Markov code in Appendix~\ref{ap:pos}, and additional comparisons against the simpler analytic models from \citet{Gieles2011} in Appendix~\ref{ap:Giel}. 

The best-fitting values we recover are similar to those previous predicted. We find that for the tidally limited cluster, $\xi_1 = 0.0142$, similar to that required by the isolated cluster, and $N_1 = 38252$, comparable to, although smaller than, that previously anticipated ($N_1 \approx 10^5$, although derived from models containing a mass function \citealt{Gieles2011}). We additionally find $z = 1.61$ (remarkably close to $z=1.5$, the value chosen for convenience in the same paper). There is evidence of some covariance between $N_1$ and $z$, which probably arises due to the manner through which these two parameters define the total lifetime of clusters. The total lifetime (see Appendix~\ref{ap:zx}) is a function of $z$, $N_1$, and $x$ such that the relationship between $z$ and $N_1$ plays a substantial part in the average rate of mass-loss throughout the lifetime. Accordingly, a variation in one is likely to cause a comparable variation in another, which will be expressed as some degeneracy.

We find that our model is able to reproduce the evolution of $N$ and $r$, and the correct behaviour in the $N - \rvrj$ plane. Figures~\ref{f:fulltides}(a) and~(b) show that the early expansion phase, transition phase, and evolution in the tidally limited regime are all well reproduced by our prescription. However, some (small) systematic offsets are visible in our prescription - typically observed around the point of transition at which the tidal field becomes significant. This is the result of two effects: first the none self-similar evolution discussed in section~\ref{s:m2}, and secondly the simplistic manner in which $\xi_{\rm tidal}$ and $\xi_1$ are combined in equation~(\ref{eq:xi3}). Nonetheless, this offset does not degrade the accuracy of this model by a significant ($>10\%$) margin, and is hence acceptable for our purpose.

It is interesting to note that in both figures~\ref{f:fulltides}(a) and~(b) $\rhrj(N)$ is not constant with decreasing $N$ in the asymptotic regime (the line upon which all our evolutionary tracks merge is not horizontal). This suggests that in this stage of evolution the density of the cluster is decreasing as stars escape. The cause of this effect is demonstrated in appendix~\ref{ap:zx} - an $N$ dependent $\xi$ for $x<1$. We find however that such an effect is not explicitly demonstrated in the evolution of $r_{\textrm h}/r_{\rm J}$ (instead, we find once again approximately constant density), although we leave the investigation of this phenomenon to a subsequent paper.

Panels (c) and (d) show the time-evolution of $N$ for our clusters. In both cases, the decrease in $N$ is adequately reproduced, although some particular series of simulations show systematic offsets. These tend to arise on account of small variations in the time taken for core collapse, and are therefore likely to be a statistical effect of our $N$-body simulations. Similar effects are also observed in the variation of $r$ (panels e and f), once again most likely on account of random perturbations in the initial distribution of stars.

\begin{figure}
\centering
    \includegraphics[width=85mm]{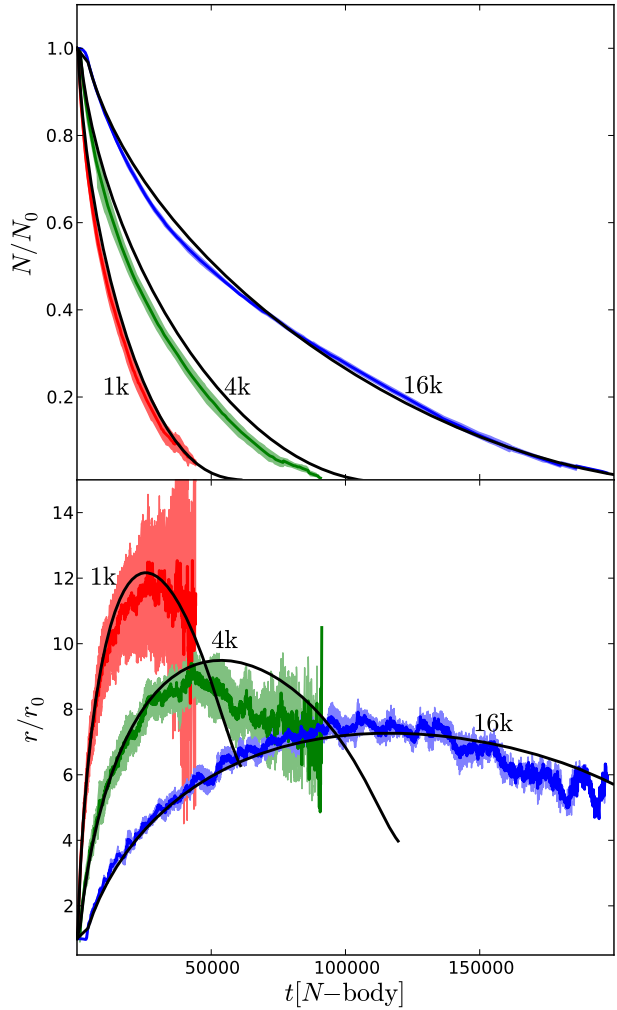}
\caption{The median values (solid lines) and standard deviations (shaded area) of our repeated simulations of $N_0=$1024, 4096, and 16384 clusters. The standard deviation is substantially greater for smaller $N_0$ simulation, as statistical noise is more significant when the number of stars is small. Accordingly, the standard deviation of our simulations increases with $t$ (and hence decreases with increasing $N$). We therefore require that the prescription describes higher $N_0$ simulations well, while the significance of our results at lower $N_0$ is reduced. In order that undue significance is not given to the early evolution (with high $N$), our limiting 10\% error is especially applicable in these regimes. The smooth (black) lines over-plotted are our individual best fits to these simulations.}
\label{f:errors} 
\end{figure}

Our prescription is generally successful in producing our $N$-body data to within the standard deviation of our series of $N$-body simulations (see figure~\ref{f:errors}). Despite this, our $N$-body data shows some overlaid noise (i.e. oscillatory behaviour), that is particularly apparent for the evolution of $r$ when compared to that of $N$. This noise becomes especially significant in later evolution (e.g. contraction), whilst expansion is comparatively smooth. We believe this effect is indicative of the ejection of stars, since every ejection of a single star or binary will be accompanied by a change (jump) in the total energy of the cluster. Accordingly, because the total energy is lower in later stages of the evolution, these jumps are more significant. These events demonstrate no particular $N$ dependency, and, as an effect corresponding to randomly occurring events, are not naturally reproduced by \textsc{EMACSS}.
 
\begin{table*}
  \centering
    \caption{Parameters defining the \textsc{EMACSS} code. Those of the left are fixed at literature values, while those on the right are recovered by a Markov Chain Monte Carlo technique. The differences in $\zeta$ are present on account of the two different definitions of $r$ used in the expression for half-mass relaxation time.} 
  \begin{tabular}{cccccccccc}
    \hline
    & \multicolumn{4}{c}{Fixed} & & \multicolumn{4}{c}{Markov Chain}\\
    &$\gamma$ & $\rhrjr$ & $x$ &$t_{\rm cc}$ & & $\zeta$ & $\xi_1$ & $N_1$ & $z$ \\
    \hline
    Isolated & 0.11 & -- & -- & 20 & & 0.105 & 0.0141 & -- & --\\ 
    Tidal & 0.11 & 0.145 & 0.75 & 20 & & 0.111 & 0.0142 & 38252 & 1.61\\ 
    \hline
  \end{tabular}
  \label{t:comp}
\end{table*}

\section{Summary and Conclusions}
\label{s:6}
 
We have developed a prescription and code to reproduce the post core-collapse evolution of the mass and radius of star clusters, built upon the work of~\citet{Henon1961,Henon1965-1}, whose combination was proposed in~\citet{Gieles2011}. The prescription is fundamentally simple, considering a cluster of equal-mass stars in the tidal field of a point-mass galaxy.

We allow an approximate form for the mass-loss during a cluster's expansion phase, followed by a smooth transition into a tidally limited phase. In this tidal phase we use the method of \citet{Gieles2008} to calculate the tidally-induced mass-loss. We additionally include the variation of the Coulomb logarithm, which has particular influence in late evolution. As a consequence, our model encompasses both expansion and contraction phases, and is applicable throughout the entire post core-collapse evolution of star clusters. Thus, we have unified, and expanded upon, the models of H{\'e}non.

We rely upon the axiom that the fractional conduction of energy from the core is constant per relaxation time. It is demonstrated that this is indeed true, for the majority of the cluster's lifetime, and that star clusters exist for the majority of their lifetime in a state of balanced evolution. From our direct measurements, we show that $\zeta \approx 0.1$, in agreement with the values predicted by H{\'e}non's models. This measured value is then used as a foundation for our prescription, which, we demonstrate, can be used to model the evolution of clusters of equal-mass stars both in isolation, and in the tidal field of a point-mass galaxy. We have shown that our prescription is generally successful, albeit with certain regions in which our prescription accuracy experiences systematic offsets.

For the isolated regime, we find that \textsc{EMACSS} can be made to approximately reproduce the simulated evolution of a star cluster for the majority of the lifetime, although also demonstrates that the evolution is not strictly self-similar. We additionally find effects associated with a gradual start to mass-loss, and the failure of the axiom of constant $\zeta$ during the final dissolution of the cluster. Meanwhile, for clusters in tidal fields, we find that the \textsc{EMACSS} code can successfully describe all stages of evolution, in terms of $N$ and $r$. There is some stochastic variation in the time taken for balanced evolution to begin in our $N$-body data, although note that if this offset is accounted for our prescription is successful.

 In conclusion, we have added a description of variable mass loss to the models of H{\'e}non, and have developed \textsc{Evolve Me A Cluster of StarS}, a numerical code that uses our description to predict the evolution of a star cluster. We have calibrated this code to recover $N$ and $r$ to within $10\%$ for the full life cycle of equal mass clusters. Although this description does not consider any of the complexity arising from full relaxation mechanisms or oscillations within the core, these effects appear to be smoothly variant over a half-mass relaxation timescale. Thus, our axiom of a constant change in energy per $t_{\textrm{rh}}$, and the dynamical mass-loss and expansion rates derived from this, are shown to be conducive to predicting the evolution of a star cluster.

\subsection{Future Development of our Code}

Our first iteration of \textsc{EMACSS} is publicly available\footnote{https://github.com/emacss}. The code predicts the evolution of mass and radius for a virialised cluster of equal mass stars, throughout it's lifetime. As the dynamical relaxation effects of the cluster are dominant throughout significant periods of a realistic cluster's life cycle, we believe this simple code will reproduce realistic star clusters to a good degree of accuracy.

The inaccuracy of our model at very low ($N \lesssim 200$) numbers of stars is acceptable, since observation of clusters on this scale is unlikely. Moreover, these small clusters are comparatively fast to model via other means. This failure does however define a lower limit for the accuracy of \textsc{EMACSS}, while our upper limit is less certain.

Despite these successes, our code remains fundamentally simple, with many physical effects neglected. Immediately apparent is the lack of a mass function; with a realistic mass function, mass segregation is likely to increase the speed of the evolution of a cluster\footnote{The most likely effect is an increase of $\zeta$, since simulations of clusters with mass functions are seen to evolve faster; \citet{Gieles2010} found that clusters with mass functions typical of globular clusters are well described when $\zeta\approx 0.2$.}. Furthermore, preferential ejection of particular (low mass) stellar types will further complicate the evolution of a cluster by changing the mean mass of stars over time. Thus, clusters with a high mass range are likely to be worse reproduced than flatter mass functions, and will require modifications to our code.

The parameters produced are limited to mass and a virial radius, which we use as an approximate representation of half-mass radius. The approximation of half-mass radius is made through the assumption that the energy form factor, $\kappa$, does not vary throughout a cluster's life cycle. It follows that by accounting for this form factor we would be able to reproduce the half-mass radius, although conversion to physical observables - such as half-light radius - would require additional consideration of the stellar mass function and exact form of density profile.

We have also considered only simple formulae to describe escape mechanisms within the cluster (e.g. our single factor form for escape rate in the isolated regime). In this way, we have overlooked several factors and mechanisms in the escape process, which may account for some of the discrepancies noted between our predictions and $N$-body simulations. A more detailed study of these mechanisms is forthcoming, with the intention of improving the fit of our model throughout the complete life cycle. 

We finally note that our prescription interprets the mass-loss due to a tidal field as an effect dependent upon a limiting Jacobi surface.  Once again, progress is forthcoming on a more precise realisation of tidal fields in which the globular cluster exists (e.g. \citealt{Lamers2010}), additional potentials and orbital eccentricity. It follows that future implementations of our code will incorporate these more versatile tidal field descriptions (\citealt{Tanikawa2010,Renaud2011}), and hence allow a greater range of situations to be studied. In conjunction with realistic (and evolving) mass functions, these improvements could significantly improve our results for realistic cluster simulations, with a potential extension into population studies of the local universe's globular clusters.

\section*{Acknowledgements}
We thank Douglas Heggie, Holger Baumgardt, Henny Lamers, Florent Renaud and Stephen McMillan for valuable discussions and suggestions related to this work. We warmly acknowledge Sverre Aarseth for consultation and support for \textsc{Nbody6}, and Keigo Nitadori for development of the GPU libraries accelerating the code. We finally acknowledge Sergey Koposov for helpful discussions on MCMC fitting, and the reviewer for helpful and constructive comments. P.A. acknowledges financial support from the Science \& Technology Facilities Council, and M.G. acknowledges financial support from the Royal Society.
\bibliography{clusters}

\begin{thebibliography}{}

\bibitem[\protect\citeauthoryear{{Aarseth}}{{Aarseth}}{1971}]{Aarseth1971}
{Aarseth} S.~J.,  1971, \apss, 14, 118

\bibitem[\protect\citeauthoryear{{Aarseth}}{{Aarseth}}{1973}]{Aarseth1974}
{Aarseth} S.~J.,  1973, Vistas in Astronomy, 15, 13

\bibitem[\protect\citeauthoryear{{Aarseth}}{{Aarseth}}{1999}]{Aarseth1999}
{Aarseth} S.~J.,  1999, \pasp, 111, 1333

\bibitem[\protect\citeauthoryear{{Aarseth}}{{Aarseth}}{2003}]{Aarseth2003}
{Aarseth} S.~J.,  2003, {Gavitational N-Body Simulations}

\bibitem[\protect\citeauthoryear{{Aarseth} \& {Heggie}}{{Aarseth} \&
  {Heggie}}{1998}]{Aarseth1998}
{Aarseth} S.~J.,  {Heggie} D.~C.,  1998, \mnras, 297, 794

\bibitem[\protect\citeauthoryear{{Ambartsumian}
  V.}{{Ambartsumian}}{1938}]{Amb1938}
{Ambartsumian} V. A.,  1938, Ann. Len. State Univ., 22, 19

\bibitem[\protect\citeauthoryear{{Baumgardt}}{{Baumgardt}}{2001}]{Baumgardt200%
1}
{Baumgardt} H.,  2001, \mnras, 325, 1323

\bibitem[\protect\citeauthoryear{{Baumgardt}, {Hut} \& {Heggie}}{{Baumgardt}
  et~al.}{2002}]{Baumgardt2002}
{Baumgardt} H.,  {Hut} P.,    {Heggie} D.~C.,  2002, \mnras, 336, 1069

\bibitem[\protect\citeauthoryear{{Baumgardt} \& {Makino}}{{Baumgardt} \&
  {Makino}}{2003}]{Baumgardt2003}
{Baumgardt} H.,  {Makino} J.,  2003, \mnras, 340, 227

\bibitem[\protect\citeauthoryear{{Chandrasekhar}}{{Chandrasekhar}}{1942}]{Chan%
dra1942}
{Chandrasekhar} S.,  1942, {Principles of stellar dynamics}

\bibitem[\protect\citeauthoryear{{Cohn}}{{Cohn}}{1979}]{Cohn1979}
{Cohn} H.,  1979, \apj, 234, 1036

\bibitem[\protect\citeauthoryear{{Drukier}, {Cohn}, {Lugger} \&
  {Yong}}{{Drukier} et~al.}{1999}]{Drukier1999}
{Drukier} G.~A.,  {Cohn} H.~N.,  {Lugger} P.~M.,    {Yong} H.,  1999, \apj,
  518, 233

\bibitem[\protect\citeauthoryear{{Fukushige} \& {Heggie}}{{Fukushige} \&
  {Heggie}}{2000}]{Fukushige2000}
{Fukushige} T.,  {Heggie} D.~C.,  2000, \mnras, 318, 753

\bibitem[\protect\citeauthoryear{{Gao}, {Goodman}, {Cohn} \& {Murphy}}{{Gao}
  et~al.}{1991}]{Gao1991}
{Gao} B.,  {Goodman} J.,  {Cohn} H.,    {Murphy} B.,  1991, \apj, 370, 567

\bibitem[\protect\citeauthoryear{{Gieles} \& {Baumgardt}}{{Gieles} \&
  {Baumgardt}}{2008}]{Gieles2008}
{Gieles} M.,  {Baumgardt} H.,  2008, \mnras, 389, L28

\bibitem[\protect\citeauthoryear{{Gieles}, {Baumgardt}, {Heggie} \&
  {Lamers}}{{Gieles} et~al.}{2010}]{Gieles2010}
{Gieles} M.,  {Baumgardt} H.,  {Heggie} D.~C.,    {Lamers} H.~J.~G.~L.~M.,
  2010, \mnras, 408, L16

\bibitem[\protect\citeauthoryear{{Gieles}, {Heggie} \& {Zhao}}{{Gieles}
  et~al.}{2011}]{Gieles2011}
{Gieles} M.,  {Heggie} D.~C.,    {Zhao} H.,  2011, \mnras, 413, 2509

\bibitem[\protect\citeauthoryear{{Giersz}}{{Giersz}}{1998}]{Giersz1998}
{Giersz} M.,  1998, \mnras, 298, 1239

\bibitem[\protect\citeauthoryear{{Giersz} \& {Heggie}}{{Giersz} \&
  {Heggie}}{1994}]{Giersz1994}
{Giersz} M.,  {Heggie} D.~C.,  1994, \mnras, 268, 257

\bibitem[\protect\citeauthoryear{{Giersz} \& {Heggie}}{{Giersz} \&
  {Heggie}}{1997}]{Giersz1997}
{Giersz} M.,  {Heggie} D.~C.,  1997, \mnras, 286, 709

\bibitem[\protect\citeauthoryear{{Goodman}}{{Goodman}}{1984}]{Goodman1984}
{Goodman} J.,  1984, \apj, 280, 298

\bibitem[\protect\citeauthoryear{{Goodman} \& {Hut}}{{Goodman} \&
  {Hut}}{1989}]{Goodman1989}
{Goodman} J.,  {Hut} P.,  1989, \nat, 339, 40

\bibitem[\protect\citeauthoryear{{Heggie} \& {Hut}}{{Heggie} \&
  {Hut}}{2003}]{Heggie2003}
{Heggie} D.,  {Hut} P.,  2003, {The Gravitational Million-Body Problem: A
  Multidisciplinary Approach to Star Cluster Dynamics}

\bibitem[\protect\citeauthoryear{{Heggie}}{{Heggie}}{1975}]{Heggie1975}
{Heggie} D.~C.,  1975, \mnras, 173, 729

\bibitem[\protect\citeauthoryear{{Heggie} \& {Aarseth}}{{Heggie} \&
  {Aarseth}}{1992}]{Heggie1992-2}
{Heggie} D.~C.,  {Aarseth} S.~J.,  1992, \mnras, 257, 513

\bibitem[\protect\citeauthoryear{{Heggie} \& {Mathieu}}{{Heggie} \&
  {Mathieu}}{1986}]{Heggie1986}
{Heggie} D.~C.,  {Mathieu} R.~D.,  1986, in {P.~Hut \& S.~L.~W.~McMillan} ed.,
  The Use of Supercomputers in Stellar Dynamics Vol.~267 of Lecture Notes in
  Physics, Berlin Springer Verlag, {Standardised Units and Time Scales}.
pp 233--+

\bibitem[\protect\citeauthoryear{{H{\'e}non}}{{H{\'e}non}}{1960}]{Henon1960}
{H{\'e}non} M.,  1960, Annales d'Astrophysique, 23, 668

\bibitem[\protect\citeauthoryear{{H{\'e}non}}{{H{\'e}non}}{1961}]{Henon1961}
{H{\'e}non} M.,  1961, Annales d'Astrophysique, 24, 369

\bibitem[\protect\citeauthoryear{{H{\'e}non}}{{H{\'e}non}}{1965}]{Henon1965-1}
{H{\'e}non} M.,  1965, Annales d'Astrophysique, 28, 62

\bibitem[\protect\citeauthoryear{{H{\'e}non}}{{H{\'e}non}}{1975}]{Henon1975}
{H{\'e}non} M.,  1975, in {A.~Hayli} ed., Dynamics of the Solar Systems Vol.~69
  of IAU Symposium, {Two Recent Developments Concerning the Monte Carlo
  Method}.
pp 133--+

\bibitem[\protect\citeauthoryear{{Hut}, {McMillan}, {Goodman}, {Mateo},
  {Phinney}, {Pryor}, {Richer}, {Verbunt} \& {Weinberg}}{{Hut}
  et~al.}{1992}]{Hut1992}
{Hut} P.,  {McMillan} S.,  {Goodman} J.,  {Mateo} M.,  {Phinney} E.~S.,
  {Pryor} C.,  {Richer} H.~B.,  {Verbunt} F.,    {Weinberg} M.,  1992, \pasp,
  104, 981

\bibitem[\protect\citeauthoryear{{Inagaki} \& {Lynden-Bell}}{{Inagaki} \&
  {Lynden-Bell}}{1983}]{Inagaki1983}
{Inagaki} S.,  {Lynden-Bell} D.,  1983, \mnras, 205, 913

\bibitem[\protect\citeauthoryear{{King}}{{King}}{1958}]{King1958}
{King} I.,  1958, \aj, 63, 109

\bibitem[\protect\citeauthoryear{{King}}{{King}}{1962}]{King1962}
{King} I.,  1962, \aj, 67, 471

\bibitem[\protect\citeauthoryear{{Lamers}, {Baumgardt} \& {Gieles}}{{Lamers}
  et~al.}{2010}]{Lamers2010}
{Lamers} H.~J.~G.~L.~M.,  {Baumgardt} H.,    {Gieles} M.,  2010, \mnras, 409,
  305

\bibitem[\protect\citeauthoryear{{Larson}}{{Larson}}{1970}]{Larson1970}
{Larson} R.~B.,  1970, \mnras, 147, 323

\bibitem[\protect\citeauthoryear{{Lee} \& {Ostriker}}{{Lee} \&
  {Ostriker}}{1987}]{Lee1987}
{Lee} H.~M.,  {Ostriker} J.~P.,  1987, \apj, 322, 123

\bibitem[\protect\citeauthoryear{Lightman \& Shapiro}{Lightman \&
  Shapiro}{1978}]{Lightman1978}
Lightman A.~P.,  Shapiro S.~L.,  1978, Rev. Mod. Phys., 50, 437

\bibitem[\protect\citeauthoryear{{Lynden-Bell} \& {Eggleton}}{{Lynden-Bell} \&
  {Eggleton}}{1980}]{LB1980}
{Lynden-Bell} D.,  {Eggleton} P.~P.,  1980, \mnras, 191, 483

\bibitem[\protect\citeauthoryear{{Lynden-Bell} \& {Wood}}{{Lynden-Bell} \&
  {Wood}}{1968}]{LB-Wood1968}
{Lynden-Bell} D.,  {Wood} R.,  1968, \mnras, 138, 495

\bibitem[\protect\citeauthoryear{{Makino}}{{Makino}}{1996}]{Makino1996}
{Makino} J.,  1996, \apj, 471, 796

\bibitem[\protect\citeauthoryear{{Makino} \& {Aarseth}}{{Makino} \&
  {Aarseth}}{1992}]{Makino1992}
{Makino} J.,  {Aarseth} S.~J.,  1992, \pasj, 44, 141

\bibitem[\protect\citeauthoryear{{Metropolis}, {Rosenbluth}, {Rosenbluth},
  {Teller} \& {Teller}}{{Metropolis} et~al.}{1953}]{Metropolis1953}
{Metropolis} N.,  {Rosenbluth} A.~W.,  {Rosenbluth} M.~N.,  {Teller} A.~H.,
  {Teller} E.,  1953, \jcp, 21, 1087

\bibitem[\protect\citeauthoryear{{Plummer}}{{Plummer}}{1911}]{Plummer1911}
{Plummer} H.~C.,  1911, \mnras, 71, 460

\bibitem[\protect\citeauthoryear{{Renaud}, {Gieles} \& {Boily}}{{Renaud}
  et~al.}{2011}]{Renaud2011}
{Renaud} F.,  {Gieles} M.,    {Boily} C.~M.,  2011, \mnras, 418, 759

\bibitem[\protect\citeauthoryear{{Spitzer} L}{{Spitzer}}{1987}]{Spitzer1987}
{Spitzer} L J.,  1987, Dynamical evolution of globular clusters.
Princeton University Press

\bibitem[\protect\citeauthoryear{{Spitzer} Jr.}{{Spitzer}}{1969}]{Spitzer1969}
{Spitzer} Jr. L.,  1969, \apjl, 158, L139+

\bibitem[\protect\citeauthoryear{{Spitzer} Jr.}{{Spitzer}}{1975}]{Spitzer1975}
{Spitzer} Jr. L.,  1975, in {A.~Hayli} ed., Dynamics of the Solar Systems
  Vol.~69 of IAU Symposium, {Dynamical theory of spherical stellar systems with
  large N}.
pp 3--26

\bibitem[\protect\citeauthoryear{{Spitzer} Jr. \& {Hart}}{{Spitzer} \&
  {Hart}}{1971}]{Spitzer1971}
{Spitzer} Jr. L.,  {Hart} M.~H.,  1971, \apj, 164, 399

\bibitem[\protect\citeauthoryear{{Spurzem}}{{Spurzem}}{1999}]{Spurzem1999}
{Spurzem} R.,  1999, Journal of Computational and Applied Mathematics, 109, 407

\bibitem[\protect\citeauthoryear{{Statler}, {Ostriker} \& {Cohn}}{{Statler}
  et~al.}{1987}]{Statler1987}
{Statler} T.~S.,  {Ostriker} J.~P.,    {Cohn} H.~N.,  1987, \apj, 316, 626

\bibitem[\protect\citeauthoryear{{Takahashi}}{{Takahashi}}{1995}]{Takahashi199%
5}
{Takahashi} K.,  1995, \pasj, 47, 561

\bibitem[\protect\citeauthoryear{{Takahashi} \& {Portegies Zwart}}{{Takahashi}
  \& {Portegies Zwart}}{2000}]{Zwart1998}
{Takahashi} K.,  {Portegies Zwart} S.~F.,  2000, \apj, 535, 759

\bibitem[\protect\citeauthoryear{{Tanikawa} \& {Fukushige}}{{Tanikawa} \&
  {Fukushige}}{2010}]{Tanikawa2010}
{Tanikawa} A.,  {Fukushige} T.,  2010, \pasj, 62, 1215

\bibitem[\protect\citeauthoryear{{von Hoerner}}{{von Hoerner}}{1957}]{VH1957}
{von Hoerner} S.,  1957, \apj, 125, 451

\end{thebibliography}
 
\appendix
\section{A general solution to $\boldsymbol{\rhrj-N}$ and $\boldsymbol{N(\lowercase{t})}$ for any $\Large{\lowercase{\boldsymbol{x}}}$ and $\Large{\boldsymbol{\lowercase{z}}}$} 
\label{ap:zx}

Although the principal objective of this study is to develop a numerical prescription for the evolution of clusters, we nonetheless explore the evolution analytically to recover (general) relationships between our fitting parameters. In particular, we explore the relationship between $x$ and $z$, as both these powers are present in $\xi_{\rm tidal}$. For the sake of a first order approximation we consider the Coulomb logarithm and $\kappa$ to be constant, and hence rewrite equation~(\ref{eq:xi2}) as 
\begin{align}
\xi = \frac{3}{5}\zeta\left(\frac{\rhrj}{\rhrjr}\right)^z\left(\frac{N}{N_1}\right)^{1-x}.
\label{eq:xi2ap}
\end{align}
We can now apply equation~(\ref{eq:xi2ap}) to equation~(\ref{eq:xi1aa}) and obtain
\begin{align}
\frac{\dr \rhrj}{\dr N} &= \frac{\rhrj}{N}\left(\frac{5}{3}-\frac{5}{3}\left(\frac{\rhrj}{\rhrjr}\right)^{-z}\left(\frac{N}{N_1}\right)^{x-1}\right). \label{eq:dRdNap}
\end{align}
This can be written as Bernoulli's differential equation, taking the form
\begin{align}
\frac{\dr \rhrj}{\dr N} + \mathcal{P}(N)\rhrj = \mathcal{Q}(N)\rhrj^{1-z} 
\end{align}
where 
\begin{align}
\mathcal{P}(N)& = -(5/3)/N\\
\mathcal{Q}(N)& = aN^{x-2} 
\end{align} 
with $a= -(5/3)\rhrj^{z}N_1^{1-x}$. We now use a variable substitution $\mathcal{U}=\rhrj^z$ to get 
\begin{align} 
\frac{1}{z}\frac{\dr \mathcal{U}}{\dr N} + \mathcal{P}(N)\rhrj^{-z} = \mathcal{Q}(N). \label{eq:dUdNap}
\end{align}
which can be solved with the correct constant of integration
\begin{align}
\mathcal{M} = \rhrjo N_0^{(5/3)z}. \label{eq:const}
\end{align}
Solving equation~(\ref{eq:dUdNap}) with~(\ref{eq:const}) we find 
\begin{align}
    \rhrj =  \frac{\rhrjr}{A^{1/z}}\left( \left[\frac{N}{N_1}\right]^{x-1}\left(1-\left[\frac{N}{N_0}\right]^{(5/3)z-x+1}\right)\right. \notag \\
    \left.+\left[\frac{\rhrjo}{\rhrjr}\right]^z\left[\frac{N}{N_0}\right]^{(5/3)z} \right)^{1/z},
  \label{eq:rn}
  \end{align}
where $A\equiv( [2/3]z+[2/5][1-x]/[2/3]z)$. For $z=3/2$ and $\rhrjo = 0$ we find equation~(A8) of \citet{Gieles2011}
\begin{align}
\rhrj  =  \frac{\rhrjr}{\left(1+\frac25(1-x)\right)^{2/3}}\left( \left[\frac{N}{N_1}\right]^{x-1}\left[1-\left[\frac{N}{N_0}\right]^{(7/2)-x}\right]\right)^{2/3}.
  \end{align}
Equation~(\ref{eq:rn}) converges in the tidal regime to the power-law relation
\begin{align}
\left(\frac{\rhrj}{\rhrjr}\right)^z \simeq \frac1A\left(\frac{N}{N_1}\right)^{x-1}.
\end{align}
Substituting this into equation~(\ref{eq:dRdNap}) we find that the logarithmic slope around $N_1$ is  
\begin{align}
\frac{\dr\ln \rhrj}{\dr\ln N} &= \frac53(1-A) \\
&= \frac{x-1}{z}.
\end{align}
Thus, the assumed values used in the model of \citet{Gieles2011} always has a logarithmic slope $(2/3)(x-1)=-1/6$. 

The value of $\xi$ converges to
\begin{align}
\xi\simeq\frac{3}{5}\frac{\zeta}{A}, \label{eq:xiap7}
\end{align}
while the relaxation time varies according to 
\begin{align}
t_{\rm rh} = t_{\rm rh,1}\left(\frac{N}{N_1}\right)\left(\frac{\rhrj}{\rhrjr}\right)^{3/2}. \label{eq:trhap7}
\end{align}
Through combination of equations~(\ref{eq:NdotN}), (\ref{eq:xiap7}) and~(\ref{eq:trhap7}), we hence find that
\begin{align}
  \dot{N}&=-\xi\frac{N}{t_{\rm rh}},\\
  &\simeq -\frac{3}{5}\zeta\frac{N_1}{t_{\rm rh,1}}\left(\frac{\rhrj}{\rhrjr}\right)^{-3/2},\\
  &\simeq -\frac{3}{5}\zeta A^{3/(2z)}\frac{N_1}{t_{\rm rh,1}}\left(\frac{N}{N_1}\right)^{3(1-x)/(2z)}. \label{eq:ndotap}
\end{align}
Combinations of $(x,z)$ can be found such that $3(1-x)/(2z)=0.25$ (i.e. $(x,z) = (1/2,3); (3/4,3/2); (5/6,1)$). These will all give the same value for $A$ as well, but smaller values for $z$ lead to slightly shorter lifetimes (equation~\ref{eq:ndotap}). This can also be understood from the asymptotic behaviour in the $\rhrj-N$ plane: smaller $z$ means smaller $\rhrj$ which means shorter relaxation time. Since all combinations of $x$ and $z$ will eventually converge to the same $\xi$, this implies the total lifetimes will be shorter.

\section{Comparison to Previous Models}
\label{ap:Giel}

\begin{figure}
  \includegraphics[width=82mm]{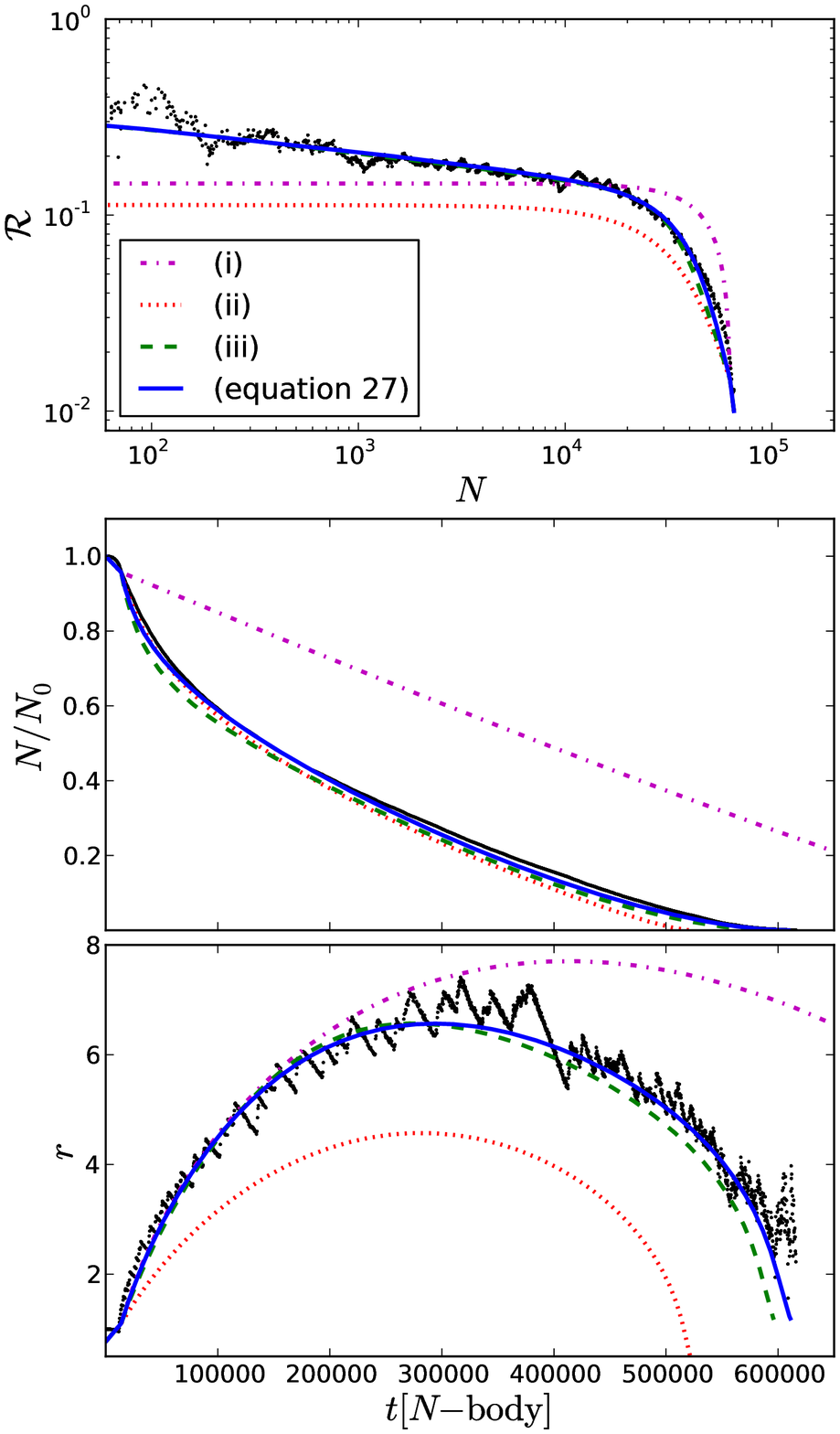}
  \caption{Different iterations of $\xi$ against $N$-body data. In case (i), the parameters used for the model are those quoted in  \citet{Gieles2011}. For the remaining cases, the models are generated using the `optimal' (best-fitting) parameters as recovered by a Markov chain. The parameters used are given in table~\ref{t:vers}, while the forms for $\xi$ are given in appendix~\ref{ap:Giel}.}
\label{f:vers}
\end{figure}

\begin{table}
  \centering
    \caption{Parameters used for the `best-fitting models' shown in figure~\ref{f:vers}. The parameters for case (i) are taken from~\citet{Gieles2011}, while the remaining models have parameters recovered by a Markov Chain Monte Carlo method. A comparable Markov Chain analysis was attempted for model (i), though failed to converge. The Markov Chain used for model (ii) also did not converge to a single value, but instead produces several `best-fitting' sets of parameters.} 
  \begin{tabular}{ccccc}
    \hline
    Parameter &\textbf{(i)} & \textbf{(ii)} & \textbf{(iii)} & \textbf{(Equation~\ref{eq:xi3})} \\
    \hline
    $\zeta$ &0.08&0.0638&0.104&105\\
    $\xi_1$ &--&0.0121&0.0085 &0.0161\\
    $\gamma$ &0.11&0.11&0.11 &0.11\\
    $\rhrjr$ &0.145&0.145&0.145 &0.145\\
    $N_1$ &--&--&54721&26689\\
    $x$ &--&--&0.566&0.731\\
    $z$ &1.5&1.5&2.67&1.75\\
    \hline
  \end{tabular}
  \label{t:vers}
\end{table}

In order to demonstrate the improvement of our new prescription when compared to previous models, we attempted comparable Markov Chain analyses for a series of models of increasing complexity. Accordingly, each model increases the number of free parameters whose values are to be recovered.
The models we attempt are as follows:
\begin{enumerate} 
\item{Model of \citet{Gieles2011}.

This model comprises the simple description of $\xi$ presented in equation (11) of \citet{Gieles2011}, namely
\begin{align}
\xi &= \frac35\zeta\left(\frac{t_{\rm cr}}{t_{\rm cr,1}}\right), \notag \\
 &= \frac35\zeta\left(\frac{\rhrj}{\rhrjr}\right)^{3/2}, \label{eq:mod1}
\end{align}
where $t_{\rm cr}$ is the (half-mass) crossing time. This model provides a description only of tidally induced mass loss, and is characterised by 4 parameters ($\zeta$, $\rhrjr$, $f_{\rm N}$ and $f_{\rm r}$).
} 
\item{ Model of \citet{Gieles2011} with constant isolated mass loss.
 
For this iteration, a second term is included in $\xi$ to account for the mass-loss in the isolated regime. Hence, $\xi$ is written as 
\begin{align}
\xi = \frac35\zeta\left(\frac{\rhrj}{\rhrjr}\right)^{z} + \xi_1,
\end{align}
and includes the additional free parameter $\xi_1$. We have also replaced the factor of $3/2$ with $z$ which we now allow to be free, and fix $\rhrjr = 0.145$ as this parameter is heavily degenerate with the combination of $z$ and $\zeta$. The effects of this variable $z$ (as opposed to a variable $\rhrjr$ are discussed in appendix~\ref{ap:zx}).
}

\item{Appendix model of \citet{Gieles2011} with constant isolated mass loss.

This form for $\xi$ is that presented in Appendix A of \citet{Gieles2011}, but with the inclusion of an additional term to describe the isolated mass loss. We hence use a modified form of \citet{Gieles2011} equation (A4), such that,
\begin{align}
\xi = \frac35\zeta\left(\frac{\rhrj}{\rhrjr}\right)^z\left(\frac{N\ln N_1}{N_1 \ln N}\right)^{1-x} + \xi_1,
\end{align}
where we have additionally included the $\ln \gamma N$ dependence of the Coulomb logarithm. We now have free parameters $\zeta$, $\xi_0$, $N_1$, $x$, $z$, $f_{\rm N}$ and $f_{\rm r}$.
}
\end{enumerate}

We use our Markov fitting code to fit the above models to a single sample of $N$-body data ($N=65536$, $\rhrjr=1/100$), and over-plot the resultant evolution for the `optimally' fitting parameters for each iteration of our prescription in figure~\ref{f:vers}. We also plot the form from our full prescription, in which $\xi$ is defined by equation~(\ref{eq:xi3}), and give our best-fitting parameters for each case in table~\ref{t:vers}. We note however that (i), the simplest version did not converge, and hence did not produce any useful best-fitting parameters. In addition, model (ii) appears to have no single value upon which the model has converged, but instead several values between which the Markov chain has oscillated.

In each case we have allowed all possible parameters to be free. Accordingly, degeneracies between certain parameter (e.g. $x$, $N_1$, $z$) are clearly visible in the joint posterior probabilities.

\section{Posterior probability distributions recovered by Markov Chains} 
\label{ap:pos}

The posterior probability densities produced by our Markov chain fitting are shown in this appendix, for a variety of situations in which we have tested our prescription. As such, the most probable values for parameters, and an estimate of uncertainty are present in the histograms, whilst degeneracies are visible in the probability density maps. Figure C1 is shows the posterior recovered for our overall fitting to isolated clusters, while figure C2 shows our overall fitting to clusters in tidal fields. The remaining figures show posteriors recovered for the simpler models described in Appendix B; figure C3 shows the posterior for fitting model (ii), figure C4 is the posterior from fitting model (iii), and figure C5 is the posterior of fitting out full prescription to a single $N_0$ = 65536, $\rvrj$ = 1/100 simulated cluster.”

\begin{figure*}
  \includegraphics[width=135mm]{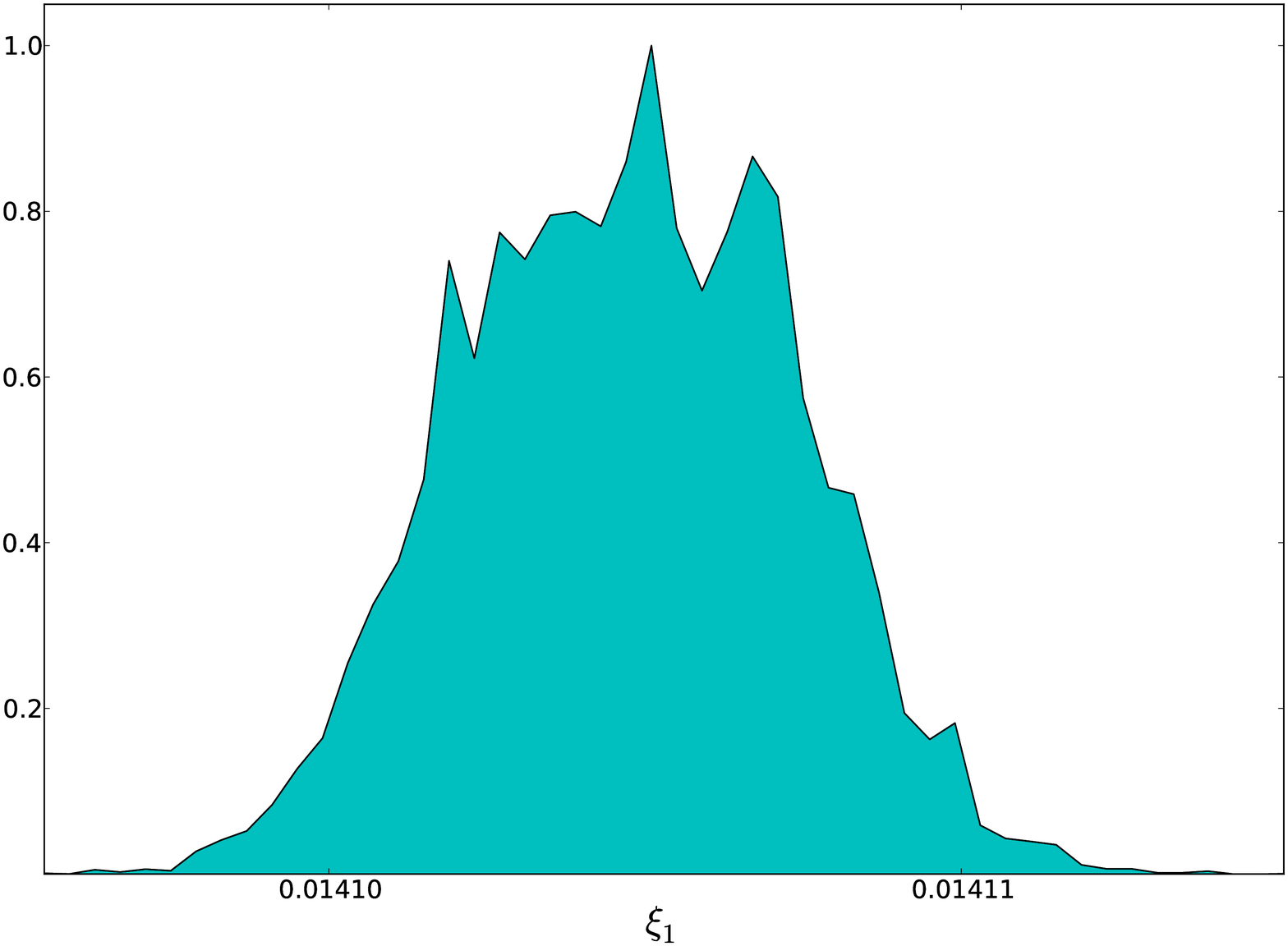}
  \caption{Histogram of the posterior probability for the $\xi_1$ of an isolated cluster. The histogram is normalised to a maximum peak value of $1$. We believe the mode value to be accurate, although the width of the posterior is likely to be underestimated on account of covariance within our data.}
\label{f:iso-01}
\end{figure*}

\begin{figure*}
  \includegraphics[width=125mm]{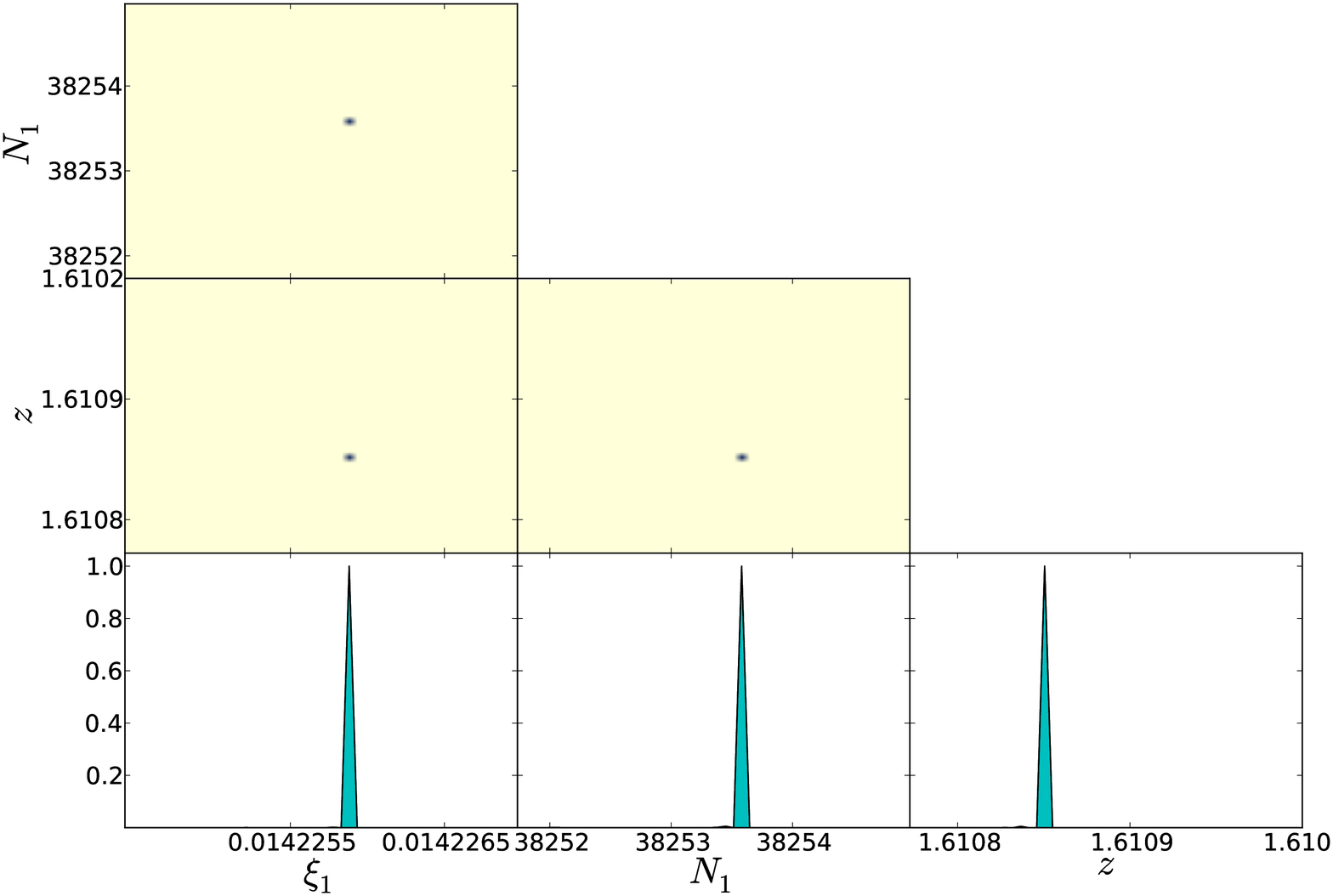}
  \caption{Joint and marginal posterior probabilities of $\xi_1$, $N_1$ and $z$ for the complete prescription, comparing the evolution of $N(t)$ and $r(t)$. The pre core collapse mass-loss and expansion, $f_{\rm N}$ and $f_{\rm r}$, are measured from $N$-body data, and $t_{\rm cc} = 20.0t_{\rm rh,0}$. The figure is generated from a single Markov chain describing all our range of $N_0$ for $\rhrjo = 1/100, 1/30$, and exhibits a very narrow peak on account of covariance within our data.}
\end{figure*}

\begin{figure*}
  \includegraphics[width=135mm]{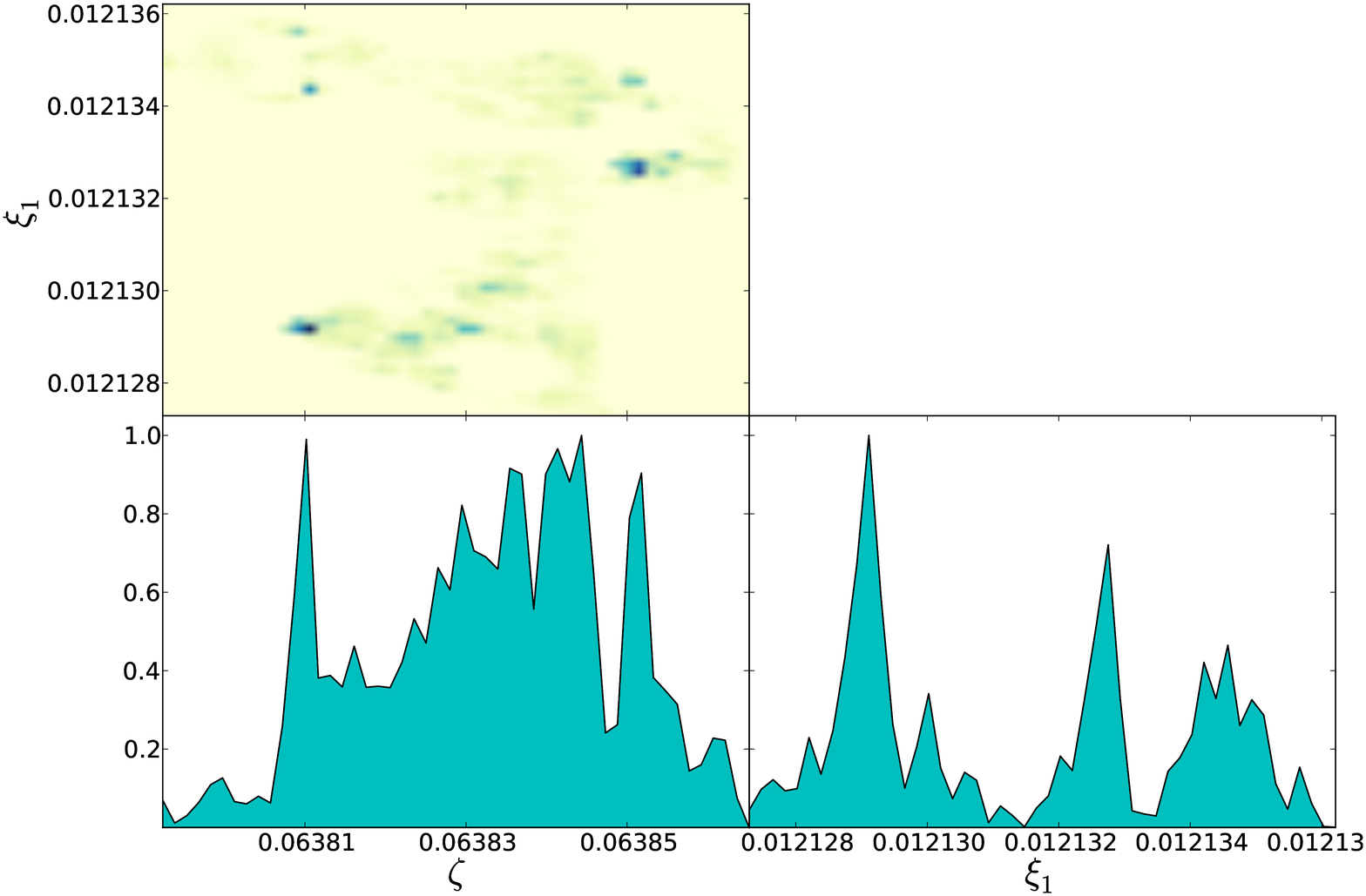}
  \caption{Joint and marginal posterior probabilities of $\zeta$ and $\xi_1$ for model (ii) in Appendix~\ref{ap:Giel}. Parameter $t_{\rm cc}$ is assigned the value $20.0t_{\rm rh,0}$, and $f_N$ and $f_r$ are measured from $N$-body data. The properties compared are the evolution of $N$ and $r$. The Markov chain has not converged in this plot, and is instead wandering between several sets of `best-fitting' values.}
\end{figure*}

\begin{figure*}
  \includegraphics[width=135mm]{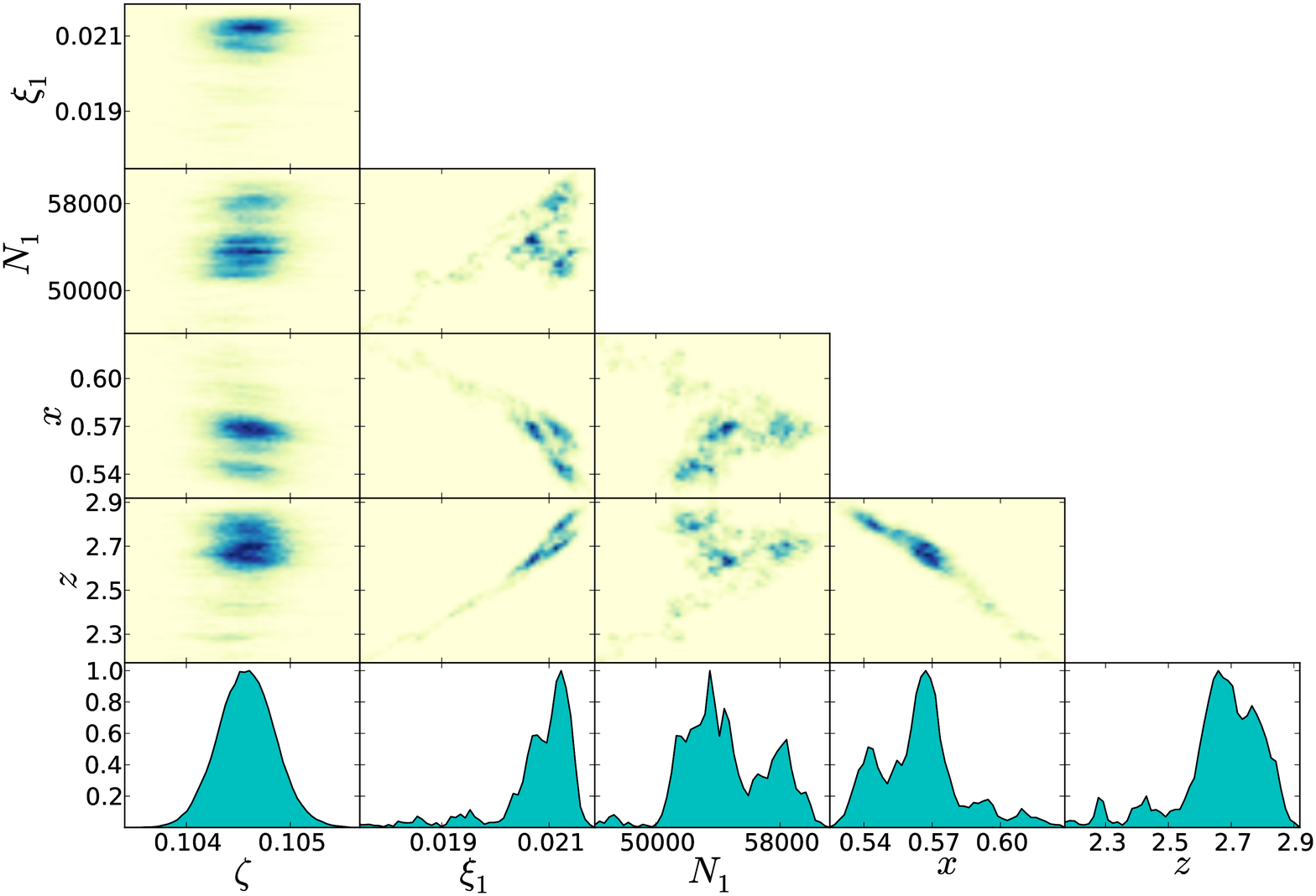}
  \caption{Joint and marginal posterior probabilities of $\zeta$, $\xi_1$,$N_1$, $x$, $z$ for model (iii) in Appendix~\ref{ap:Giel}. Parameter $t_{\rm cc}$ is assigned the value $20.0t_{\rm rh,0}$, and $f_N$ and $f_r$ are measured from $N$-body data. The properties compared are the evolution of $N$ and $r$.}
\end{figure*}

\begin{figure*}
  \includegraphics[width=135mm]{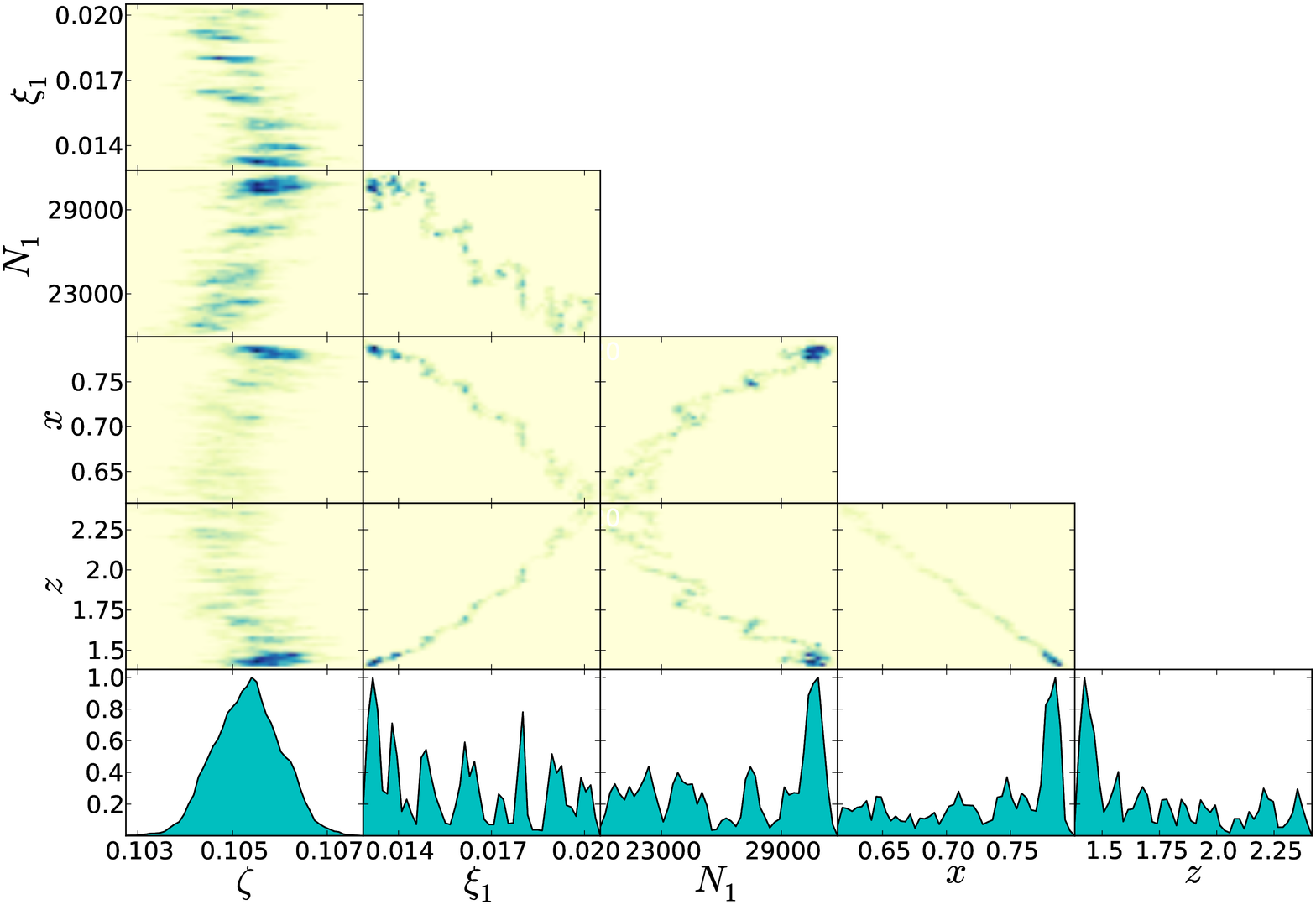}
  \caption{Joint and marginal posterior probabilities of $\zeta$, $\xi_1$, $N_1$, $x$, and $z$ for the complete prescription. Parameter $t_{\rm cc}$ is assigned the value $20.0t_{\rm rh,0}$, and $f_N$ and $f_r$ are measured directly from $N$-body data. The properties compared are the evolution of $N$ and $r$. Degeneracies between $x$, $N_1$, $\xi_1$ and $z$ are clearly visible.}
\label{f:apf}
\end{figure*}

\end{document}